\pdfoutput=1

\documentclass{emulateapj}
\slugcomment{{\sc Accepted to ApJS:} December 5, 2012} 

\shortauthors{Walkowicz et al.}  
\usepackage[caption=false]{subfig}

\begin{document}

\title{The Information Content in Analytic Spot Models of Broadband Precision Lightcurves}

\author{
Lucianne M. Walkowicz\altaffilmark{1},
Gibor S. Basri\altaffilmark{2},
Jeff A. Valenti\altaffilmark{3}
}

\altaffiltext{1}{Department of Astrophysical Sciences, Princeton University, Peyton Hall, 4 Ivy Lane, Princeton NJ 08534}
\altaffiltext{2}{Astronomy Department, University of California at Berkeley, Hearst Field Annex, Berkeley, CA 94720}
\altaffiltext{3}{Space Telescope Science Institute, 3700 San Martin Dr., Baltimore MD 21218}

\begin{abstract}
We present the results of numerical experiments to assess degeneracies in lightcurve models of starspots. Using synthetic lightcurves generated with the Cheetah starspot modeling code, we explore the extent to which photometric light curves constrain spot model parameters, including spot latitudes and stellar inclination. We also investigate the effects of spot parameters and differential rotation on one's ability to correctly recover rotation periods and differential rotation in the Kepler lightcurves. We confirm that in the absence of additional constraints on the stellar inclination, such as spectroscopic measurements of {\em v sin i} or occultations of starspots by planetary transits, the spot latitude and stellar inclination are difficult to determine uniquely from the photometry alone. We find that for models with no differential rotation, spots that appear on opposite hemispheres of the star may cause one to interpret the rotation period to be half of the true period. When differential rotation is included, the changing longitude separation between spots breaks the symmetry of the hemispheres and the correct rotation period is more likely to be found. The dominant period found via periodogram analysis is typically that of the largest spot. Even when multiple spots with periods representative of the star's differential rotation exist, if one spot dominates the lightcurve the signal of differential rotation may not be detectable from the periodogram alone. Starspot modeling is applicable to stars with a wider range of rotation rates than other surface imaging techniques (such as Doppler imaging), allows subtle signatures of differential rotation to be measured, and may provide valuable information on the distribution of stellar spots. However, given the inherent degeneracies and uncertainty present in starspot models, caution should be exercised in their interpretation.

\end{abstract}


\section{Introduction}
\label{sec:intro}

Starspots are a ubiquitous and often dramatic manifestation of stellar activity. While their study dates back to Galileo's  observations of our own lightly freckled Sun, modern improvement in the precision of ground-based photometry, along with sophisticated spectroscopic techniques such as Doppler imaging, have allowed us to observe spots on stars both like our Sun and considerably different from it \citep[see the review of][]{2009A&ARv..17..251S}.  However, these analyses have been performed on a relatively small number of stars, and don't completely sample the full range of activity for stars like our Sun and cooler. The advent of high precision space-borne photometric missions such as \textit{Microvariability and Oscillations of Stars} \citep[MOST,][]{2003PASP..115.1023W}, \textit{Convection, Rotation and planetary Transit}
\citep[CoRoT,][]{2009A&A...506..411A}, and most recently \textit{Kepler} \citep{2010ApJ...713L..79K}, provide a new window into understanding starspots though long term photometric monitoring of a range of different spectral types and magnetic activity. 

Starspots are thought to be surface manifestations of tubes of magnetic flux that form at the shearing interface between the convective and radiative zones (or ``tachocline''), which rise buoyantly through the convective envelope, become tangled by shearing motions, and protrude from the stellar surface. Starspots appear as the foot points of these loops, where suppression of convection by strong magnetic fields prevents hotter material from rising and causes the spot to remain cooler (and therefore darker) than the rest of the photosphere. Though the exact physical details remain elusive, spot coverage and the distribution of spots on the stellar surface are a function of the strength and geometry of the magnetic field \citep{2011PEPI..187...78H}. The generation of the magnetic field is intimately linked to stellar rotation, and indeed the modulation due to starspots is often used to determine the rotation rate of stars. However, stars rotate differentially, so ``the'' stellar rotation rate is not really best characterized by a single value but rather by a range of rotation rates corresponding to various latitudes, and depending on the amount of differential rotation at the stellar surface. On top of modulating the lightcurve as they rotate into and out of view, spots and active regions display additional time variable behavior as they evolve-- emerging, growing, shrinking, and vanishing. 

Due to the Sun's proximity, there exists a wealth of data on the surface manifestations of its magnetic activity. Unsurprisingly, many models of the solar-like dynamo are therefore specifically tuned towards reproducing observations of the Sun, e.g. the preferential appearance of pairs of spots of opposite polarity, which appear at preferred latitudes that drift over the course of a solar cycle \citep{2011PEPI..187...78H}. Modern observations of spots on other main sequence stars have therefore proved surprising in many cases, revealing spots with huge areal coverage, and some spots even located at the stellar pole \citep[a feature never observed on our star;][]{2007MNRAS.377.1488H,2008MNRAS.384...77M, 2009A&ARv..17..251S}. However, the Sun remains the only main sequence star for which detailed surface imaging is possible, and so in this sense we study our own star very differently than we study others. 

To bridge the solar and stellar regimes, astronomers have long turned to phenomenological models to reproduce the distribution of surface features that account for changes in the integrated light of a given star. These models are generally divided into two categories: surface integration techniques, which divide the stellar surface into a pixelated grid and integrate over the contributions of all surface elements \citep[such as the Maximum Entropy models employed in the detailed modeling of CoRoT-7;][]{2010A&A...520A..53L}, and analytical models, which use equations to calculate the loss of light due to spots on the stellar surface \citep[such as][and the one presented here]{1977Ap&SS..48..207B,1979ApJ...227..907E,1987ApJ...320..756D,2006ApJ...648..607C,2009A&A...506..245M}. Analytic models  have an advantage over surface integration in that the results from surface integration are sensitive to the resolution of the model star (i.e. the number of surface elements or ``pixels''), but surface integration is capable of modeling spots of any shape while analytical models typically require the spot shape to be simple (i.e. circular spots). Analytic models also typically have very few free parameters, whereas more complex models tend to have many; therefore, although time-variable surface integration models can attain good fits across an entire lightcurve, in some sense the inherent degeneracies and uncertainties are concealed within the number of ``knobs'' available to tweak the model. Analytic models have a particular benefit for the  exploration of parameter degeneracies presented in this paper, in that certain parameters (e.g. spot location, number of spots, spot size, etc) are  well defined in analytic models, but not so well defined in the case of a continuous surface distribution. Therefore, it is difficult to use surface integration models to explore the relative effects of any of these parameters individually. In these two methods are also two philosophical approaches to spot modeling: detailed modeling that may provide ground truth for a single star, or simpler approaches from which overall trends in spot coverage and location can be determined. Of course, these two approaches can dovetail nicely; in the case of the \textit{Kepler} dataset, the very large number of observed stars (160,000+) offers an opportunity to obtain statistical knowledge of the spot coverage and distribution through simpler analysis of a large number of stars, while the precision of the data also allows detailed modeling of individual cases. 

Photometric spot models are complementary to (and in some cases possess advantages over) spectroscopic mapping techniques, such as Zeeman Doppler imaging (ZDI) and Doppler imaging. ZDI is only suited to rapidly rotating stars for which the magnetic field is strong enough to produce detectable circular polarization signal \citep[see for example][among others]{1997A&A...326.1135D,2008MNRAS.390..545D,2008MNRAS.384...77M}. ZDI does possess the advantage that it yields a spectroscopic constraint on the stellar inclination by measurement of {\em v sin i}, which is a difficult quantity to determine from photometry alone (as we discuss later). As ZDI is effectively insensitive to cool spots, due to the polarization signal being dominated by bright magnetic features (e.g. faculae and network), the maps provided by ZDI are not directly comparable to maps derived from spot models. Doppler imaging in unpolarized light (via mapping spectral line asymmetries to spatial position on the star) can constrain cool spots like the models we discuss here, but requires additional spectroscopic followup for individual targets \citep{1983PASP...95..565V}. Methods like the spot modeling discussed in this paper can exploit the large sample of precise \textit{Kepler}  lightcurves as a powerful tool for understanding magnetic activity in stars. 

Ultimately, we would like to relate the photometric variability we observe to the physical properties of the star, such as spot coverage, locations, emergence timescales, etc. However, their interpretation via spot modeling requires careful analysis and an understanding of what one can and cannot know with confidence from the data itself. We therefore investigate whether metrics, developed from the morphology of the lightcurves themselves, can be related to physical characteristics of the spot distribution. Finding a set of morphological metrics would then allow constraints to be placed on the spot model parameters prior to fitting, potentially breaking some of the pervasive degeneracies inherent in spot model fitting. Examples of morphological metrics have been useful in previous examinations of the Kepler data; for example, in \citet{2011AJ....141...20B}, we showed that ``zerocrossings'', or the number of times a lightcurve crossed its median value, could be used to represent a characteristic timescale of variability. When related to the distribution of peaks in the Lomb-Scargle periodogram, this purely morphological metric provided a means of distinguishing dwarfs and giants on the basis of their lightcurve morphology alone. It is this kind of practical metric we seek in this paper. 

In this paper, we use a simple analytic starspot model that we are currently developing, Cheetah\footnote{So named because it's fast with spots.} to create a suite of synthetic lightcurves whose underlying parameters are known. Our eventual intent is to apply Cheetah to produce simple models for a large number of Kepler lightcurves, at which time a complete treatment of the fitting behavior and performance quality of the code will be addressed. Here, we take the opportunity to investigate the question that underlies any spot modeling effort, ``what is the information content of the chosen diagnostic?'', by asking what metrics, based on the the lightcurve morphology alone, may be related to physical parameters that were the known inputs into the synthetic curves. 

In the following section, we discuss the model parameters and generation of the suite of synthetic lightcurves. We then discuss the library of model lightcurves created, and use these models to analyze the underlying degeneracies inherent in starspot modeling. 

\section{Description of the Model}
\label{sec:modeldescrip}

Cheetah is an analytical model that uses the formulation of \citet{1994ApJ...420..373E} to calculate the modulation of a lightcurve due to starspots-- this formulation is equivalent to other analytical models \citep[e.g.][]{1977Ap&SS..48..207B,1987ApJ...320..756D}; use of these particular formulae comes down to coding preference. The main parameters of the program are the linear and quadratic limb darkening coefficients, stellar inclination, spot locations and sizes, and the intensity ratio of the spots to the stellar photosphere. In all cases, the unspotted ``pristine'' luminosity of the star is unknown, so we set the continua of our lightcurves to 1 at their maximum-- this choice implies that inferred spot coverage will miss any non-varying component that might exist (e.g., symmetric polar spots). Unfortunately, this uncertainty plagues all starspot modeling efforts, and was shown by  \citet{1997A&A...323..801K} to be a major error source in fitting for the true spot parameters. As the current work uses only a metric-based approach based on model lightcurves, rather than a treatment of the code fitting performance, we do not explore this uncertainty further here.

\subsection{Model Parameters}

Cheetah uses circular spots, described in the model by their radii, contrast, and their location in stellar latitude and longitude. Although the spots are circular, it is possible to build more elaborate (and likely more realistic) shapes through groupings of many small spots. The number of spots is not limited, but the program seeks to use the minimum number of spots to produce a good fit. Although we refer to them as spots for simplicity's sake, the spots in a simple model such as the one presented here should be thought of as a representation of spot groups (or ``active regions'') as opposed to true single spots themselves. It has been noted by many others undertaking an effort such as this that lightcurves may be as equally well fit by a few large spots as by many small spots. To add many spots to a model is to ascribe a significance to the individual spot locations and sizes that may not be justified.

There is currently scant data regarding the distribution of spot temperature as a function of spectral type, but indications are that the temperature difference between spots and their surrounding photospheres varies with stellar temperature, such that the coolest star measured (M4V, T$_{eff}$ $\sim$ 3300K) has $\Delta$T $\sim$ 200 K, while the hottest star measured (G0V, T$_{eff}$ $\sim$ 5800K) has $\Delta$T $\sim$ 2000 K \citep{2005LRSP....2....8B}. Using Cheetah it is possible to either fit or fix the contrast ratio between the spot and photosphere, however with only the broad white-light bandpass lightcurves from Kepler, it is impossible to constrain the temperature (or temperature distribution) of the spots. In practice, we therefore fix the spot-to-photosphere intensity ratio at the rough solar value, 0.67 \citep{2003A&A...403.1135L}. There is some degeneracy between the spot contrast ratio and the spot area coverage, as similar lightcurves result from smaller, darker spots as from larger spots with lower contrast \citep{1981ApJ...250..327V}. However, while these two situations might result in similar depths in the spot features, they do not produce an exactly similar shape lightcurve as larger and smaller spots create different gradients as they rotate into view around the limb. Detailed fitting to the lightcurve can therefore aid in understanding the spot coverage for a given contrast ratio even if this contrast is not the ground truth. 

In principle, we could hold more closely to the solar analogy and attempt to account for umbral and penumbral spot contributions by allowing the spot contrast to vary for each spot. However, we favor a uniform spot contrast in our models, as there are currently no observational constraints for umbral/penumbral structure in starspots \citep{1994ApJ...420..373E,1988AJ.....96..741L,1987ApJ...320..756D}. Bright regions are also neglected in the current formulation. Unlike spots, bright faculae are most visible on the stellar limb, whereas spots have their maximum effect on the lightcurve at disk center. In our Sun, the faculae may be the dominant contributor to the lightcurve morphology \citep{2003A&A...403.1135L}, while the lightcurves of more active stars seem to be dominated by cool spots \citep{2007ApJS..171..260L}. It is particularly difficult to separate the effect of plage from spots in the analysis of light curves where the pristine continuum is unknown-- while plage have their main effect as the spots are appearing or disappearing around the stellar limbs, in the case of multiple spots these effects will be hard to recognize, and in the single spot case it may simply cause one to estimate the continuum as being higher \citep[see, e.g.,][]{1992SoPh..142..197P}. Many stars in the Kepler dataset don't present an urgent need to invoke faculae in the models, so we err on the side of model simplicity and neglect bright surface features for now. However, to truly join our knowledge of the solar and stellar regimes where faculae may dominate the effect of cool spots, we anticipate that faculae will have to be included in a future version of the model.

As in the case of modeling planetary transits, it is crucial to include the effects of stellar limb darkening. Cheetah employs the quadratic limb darkening law put forth by \citet{2004A&A...428.1001C}:
\begin{equation}
I(\mu) = I(0) \times (1 - LD1\times(1-\mu) - LD2 \times(1-\mu^2))
\end{equation}
where $\mu=cos(\theta)$ and $\theta$ is the viewing angle, LD1 is the linear term in the limb darkening law, and LD2 is the quadratic term. New limb darkening coefficients, intended to conform with the \citet{2004A&A...428.1001C} formulation, have been calculated for a grid of ATLAS LTE stellar atmospheres using the Kepler bandpass by \citet{2010A&A...510A..21S}. For the models we discuss below, we chose an arbitrary LD1 = 0.45 and LD2 = 0.3. Since we only compare between these synthetic lightcurves, the exact value of the limb darkening coefficients is not important in this case.

\subsection{Synthetic Lightcurves}

In addition to helping translate photometric data to the physical properties of the star, spot models can also provide a synthetic dataset for which the physical answers are known a priori. These synthetic lightcurves are a powerful tool in understanding the inherent uncertainties and degeneracies in the spot modeling effort. In the following sections, we explore what constraints the photometry itself places on the likely spot properties and distribution, in an attempt to understand what information can be recovered, and with what certainty, from the Kepler data. 

We generated several suites of synthetic lightcurves having either one or two spots, where single spot models had a single rotation period of 10 days, and two-spot models could either be in solid-body rotation (both spots having a period of 10 days) or differentially rotating (one spot having a 10 day rotation period, the other a 12 day rotation period). In all cases, the lightcurves have a cadence of 30 minutes (roughly comparable to the Kepler ``long cadence'' data) and a duration of 90 days (comparable to the length of an individual Kepler lightcurve for a single observing ``quarter'', or three months' time). The single spot lightcurves consist of stars having a range of inclinations from 10$^\circ$ to 80$^\circ$ (in increments of 10) and a spot of either 2$^\circ$, 4$^\circ$, or 8$^\circ$ radius located at any latitude between $-$80$^\circ$ and 80$^\circ$ (also in increments of 10). As longitude affects only the phase where one sees a spot appear, all single spots were placed at longitude of 180$^\circ$. 

We created four sets of two-spot models: a set at 90$^\circ$ (equator-on) inclination, with and without differential rotation, and a set at 45$^\circ$ inclination, also with and without differential rotation. The two spots could have latitudes of -60$^\circ$, -30$^\circ$, 0$^\circ$, 30$^\circ$, or 60$^\circ$, and any longitude in increments of 10$^\circ$. Each spot could be 2$^\circ$, 4$^\circ$, or 8$^\circ$ in radius. The spots were placed on the star using random combinations from the available locations and sizes. Differential rotation was not required to be solar-like (i.e. lower latitudes were not required to rotate faster). In all our models, spots were fixed at a contrast of 0.67 (as previously discussed). We will therefore discuss spots in terms of their size or projected size, but readers should bear in mind that the spot contrast and the inferred size of a given spot are related and even degenerate quantities. 

\section{The Information Content of Lightcurves}
\label{sec:results}

\begin{figure*}
\begin{center}
\subfloat[][]{\label{tutorial:p-a}\includegraphics[width=0.45\textwidth]{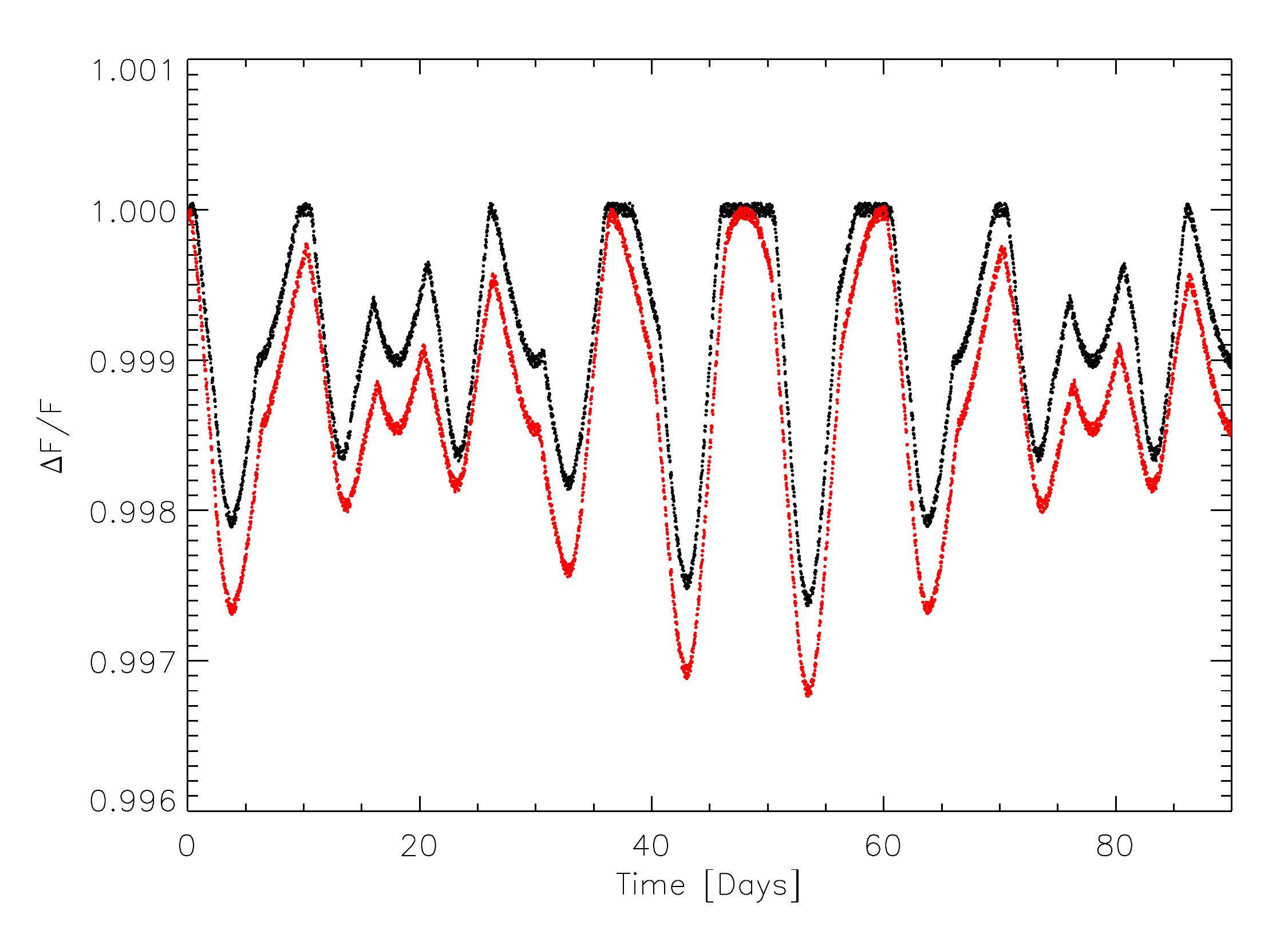}} 
\subfloat[][]{\label{tutorial:p-b}\includegraphics[width=0.45\textwidth]{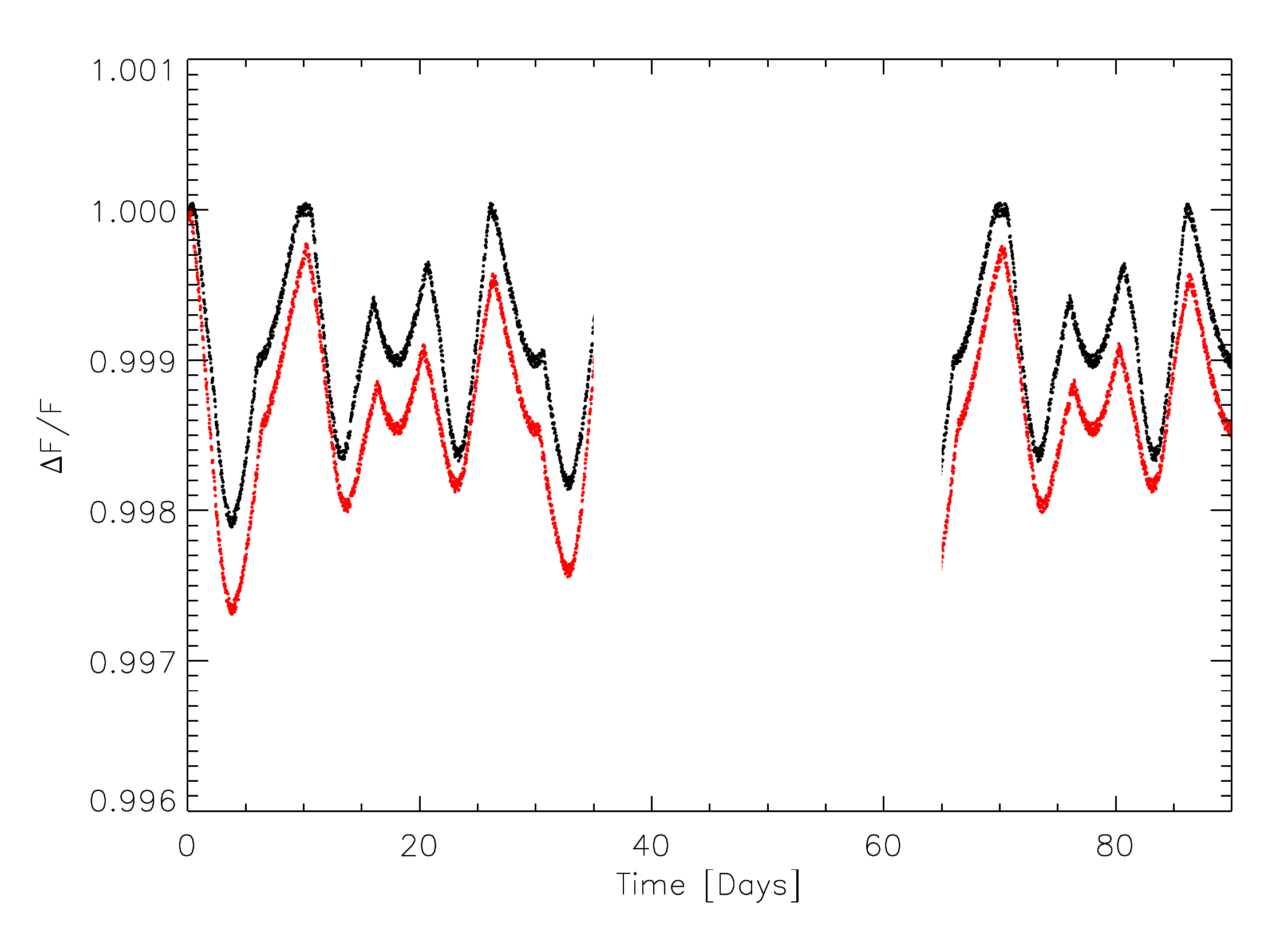}} \\
\subfloat[][]{\label{tutorial:p-c}\includegraphics[width=0.45\textwidth]{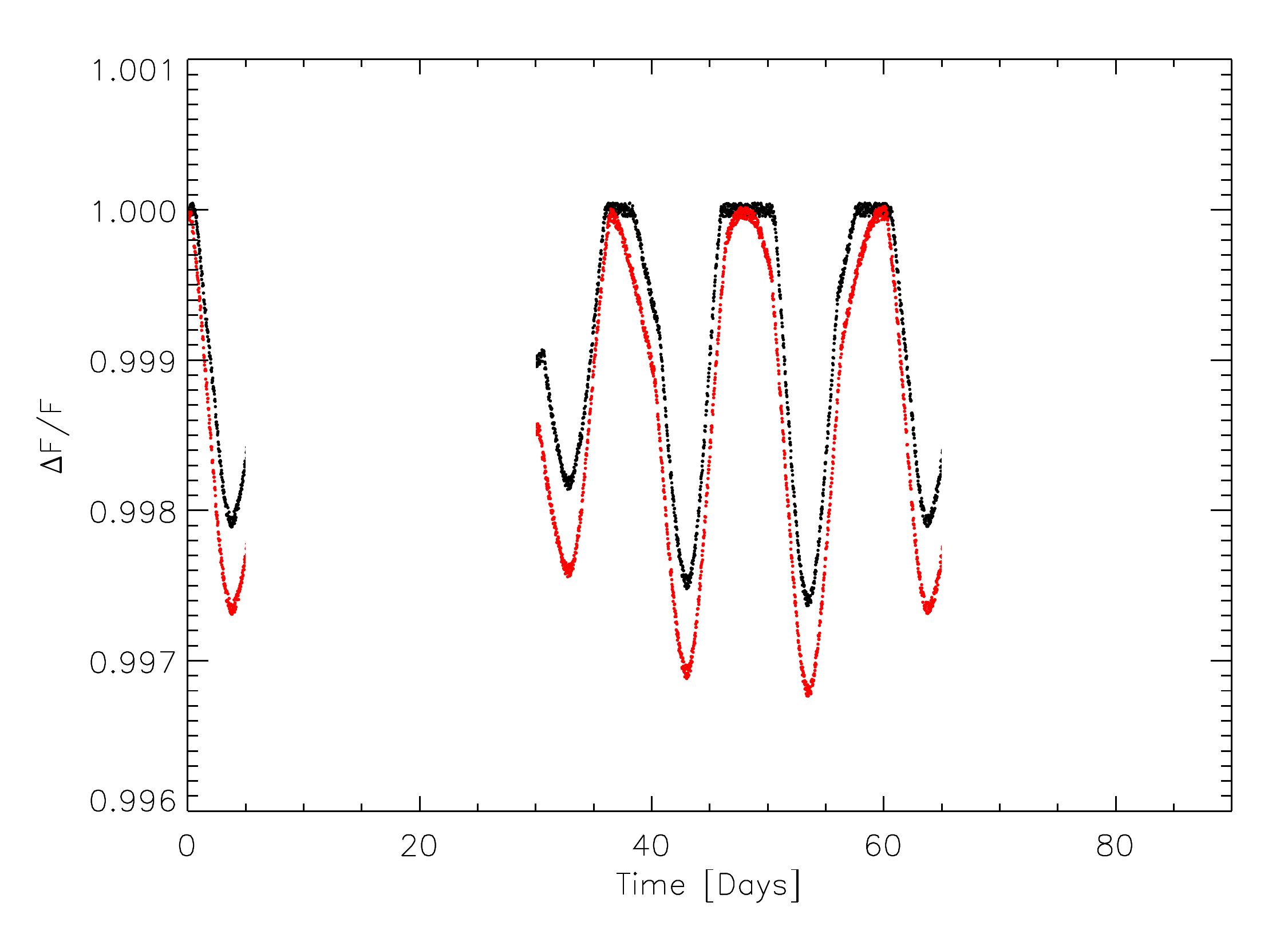}} 
\end{center}
\caption{The spot morphology at the time of observation affects one's ability to correctly deduce spot parameters and stellar properties from the photometry. Panel A shows two lightcurves having exactly the same spot locations, sizes and contrast, but different stellar inclinations (80 degrees for the black lightcurve, 60 degrees for the red lightcurve). This particular model has differential rotation, causing the morphology of the spot features to change with time. In the following two panels, B and C, we show these same lightcurves with parts blocked out to illustrate the effect of having viewed these two lightcurves at different times. In Panel B, the fact that a spot is almost always visible results in a lightcurve with no ``continuum'' or unspotted level, and therefore these two lightcurves would be difficult to distinguish from one another. In Panel C, one can clearly see a difference in the ingress and egress shape of the features for the two inclinations, but the depth of the features might lead one to infer a darker spot for the deeper features of the red 60 degree inclination lightcurve, rather than this effect being solely due to the projection of the spot size. }
\label{tutorial}
\end{figure*}

In the discussion below, we will delve into how specific parameters affect models and the ability to recover those parameters from the lightcurves themselves. Before we do so, however, we pause here to provide a visual tutorial on the inherent difficulty of modeling stellar spots. In Figure \ref{tutorial:p-a}, we compare two two-spot lightcurves. From the full 90 day lightcurves in panel \ref{tutorial:p-a}, it is evident that these two lightcurves, while similar, have notable differences. In panel \ref{tutorial:p-b}, we have blocked out the central part of the lightcurve, where a true unspotted continuum is visible. Keeping in mind that in the case of actual data, the unspotted continuum level is not known a priori, one might be hard-pressed to tell the difference between the red and black curves if they were normalized to the same maximum value. At bottom, panel \ref{tutorial:p-c} again shows these two lightcurves, but now with the opposite portions blocked out. Here the shapes of the ingress and egress of the features are markedly different, with the black lightcurve having sections of truly unspotted, flat continuum. The depth of the features appears to be different, though, and so one might be tempted to invoke a darker spot to account for this effect. However, the two lightcurves shown here have exactly the same spot locations, sizes, spot contrast, and differential rotation$-$ their only difference is that one has a stellar inclination of 80$^\circ$ (black) and the other a lower inclination of 60$^\circ$ (red). Throughout the entire following discussion, it bears remembering that the conclusions one draws on a given portion of a lightcurve may require revision in the presence of additional data (such as the resolution of closely-spaced frequencies as the duration of the time series increases; see end of Section \ref{sec:perrecover}). Indeed, one of the greatest benefits provided by the {\em Kepler} data is the duration of the observations, especially in the presence of differential rotation, where the beat period between different spot periods can be long, and so constraints on spot parameter improve with time.

\subsection{Latitude/Inclination Degeneracy}
\label{sec:latincldegen}

\begin{figure*}
\label{ingress}
\subfloat[][]{\label{ingress:my-a}\includegraphics[width=0.45\textwidth]{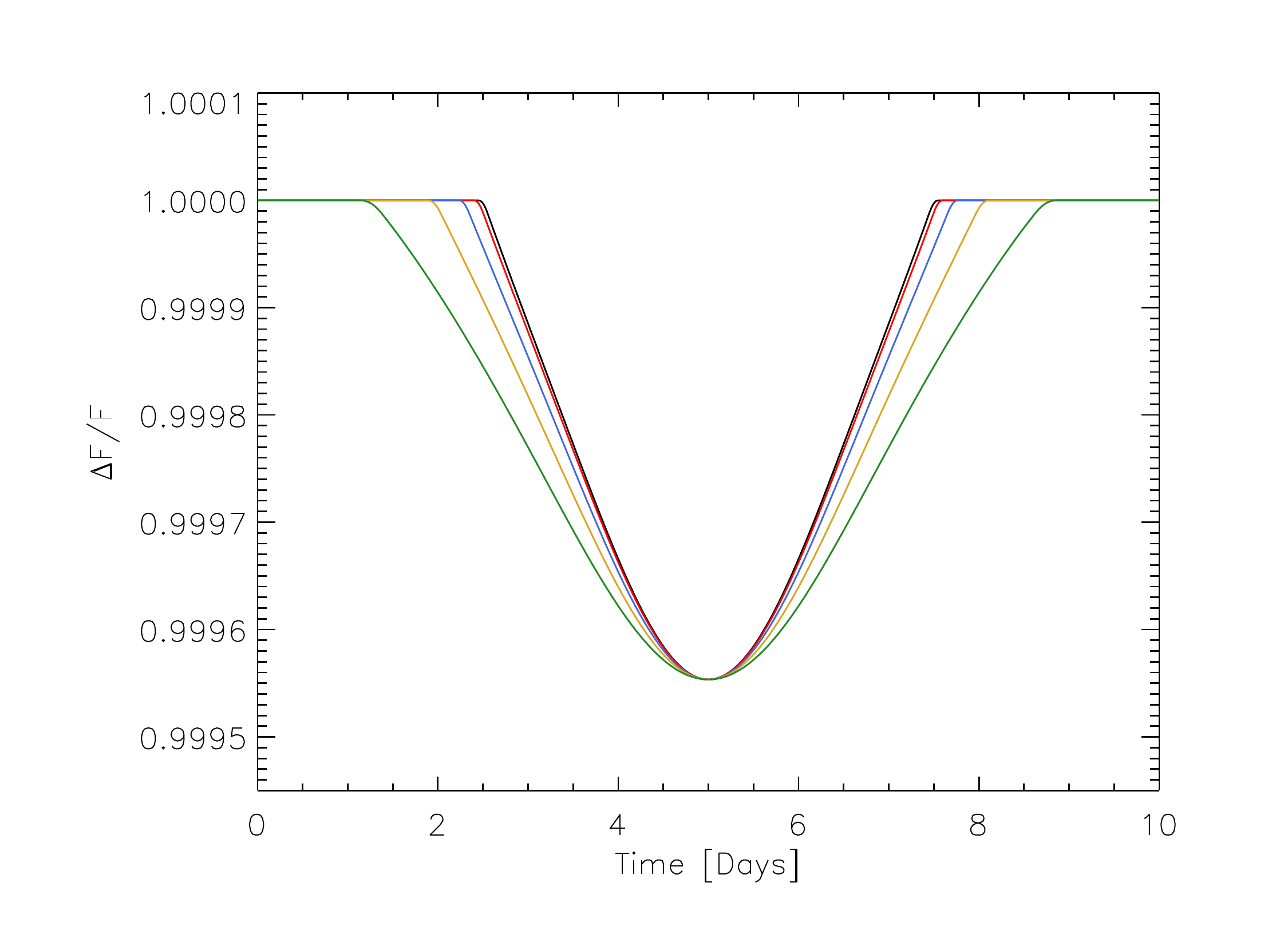}} 
\subfloat[][]{\label{ingress:my-b}\includegraphics[width=0.45\textwidth]{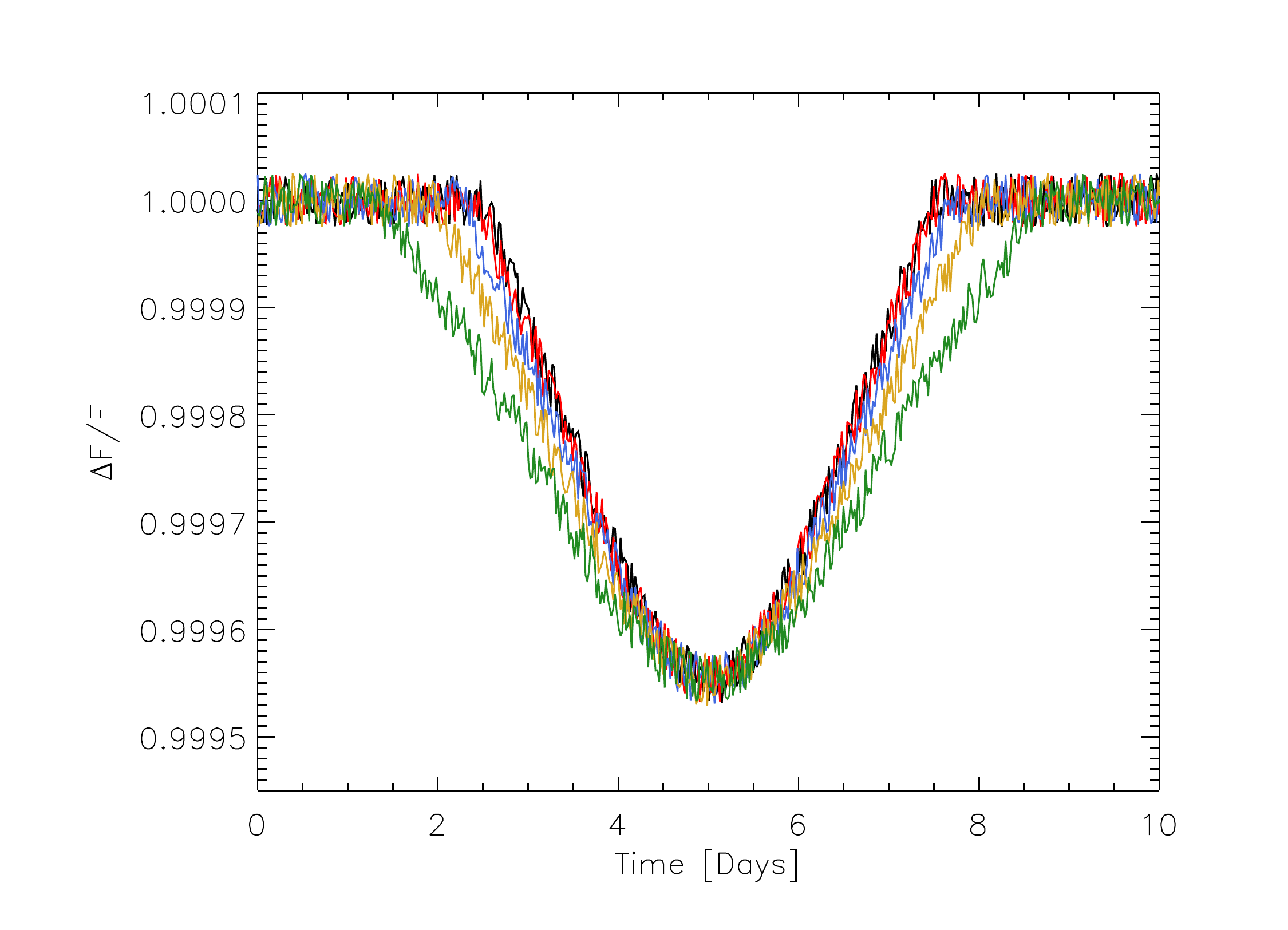}} \\
\subfloat[][]{\label{ingress:my-c}\includegraphics[width=0.45\textwidth]{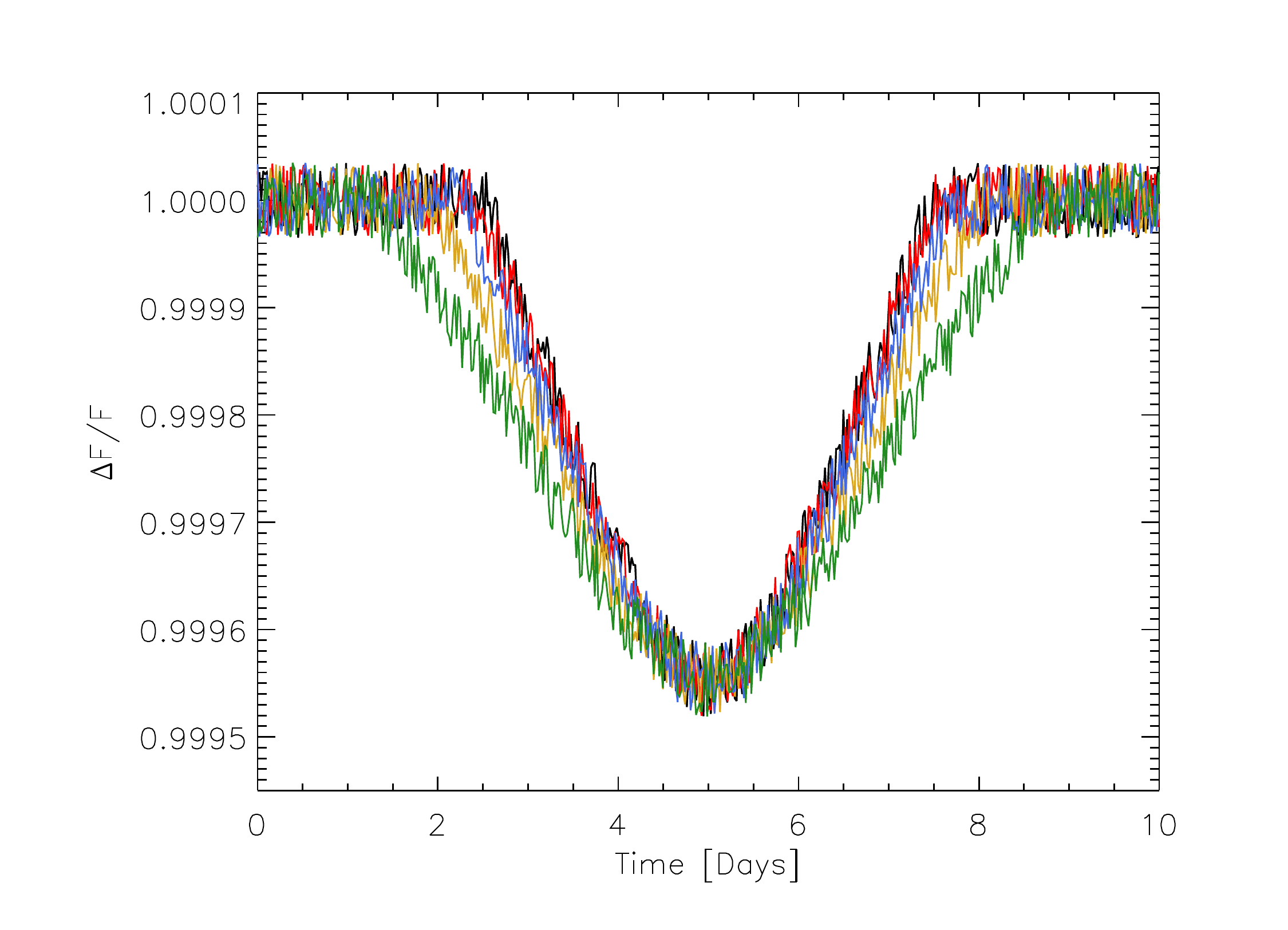}} 
\subfloat[][]{\label{ingress:my-d}\includegraphics[width=0.45\textwidth]{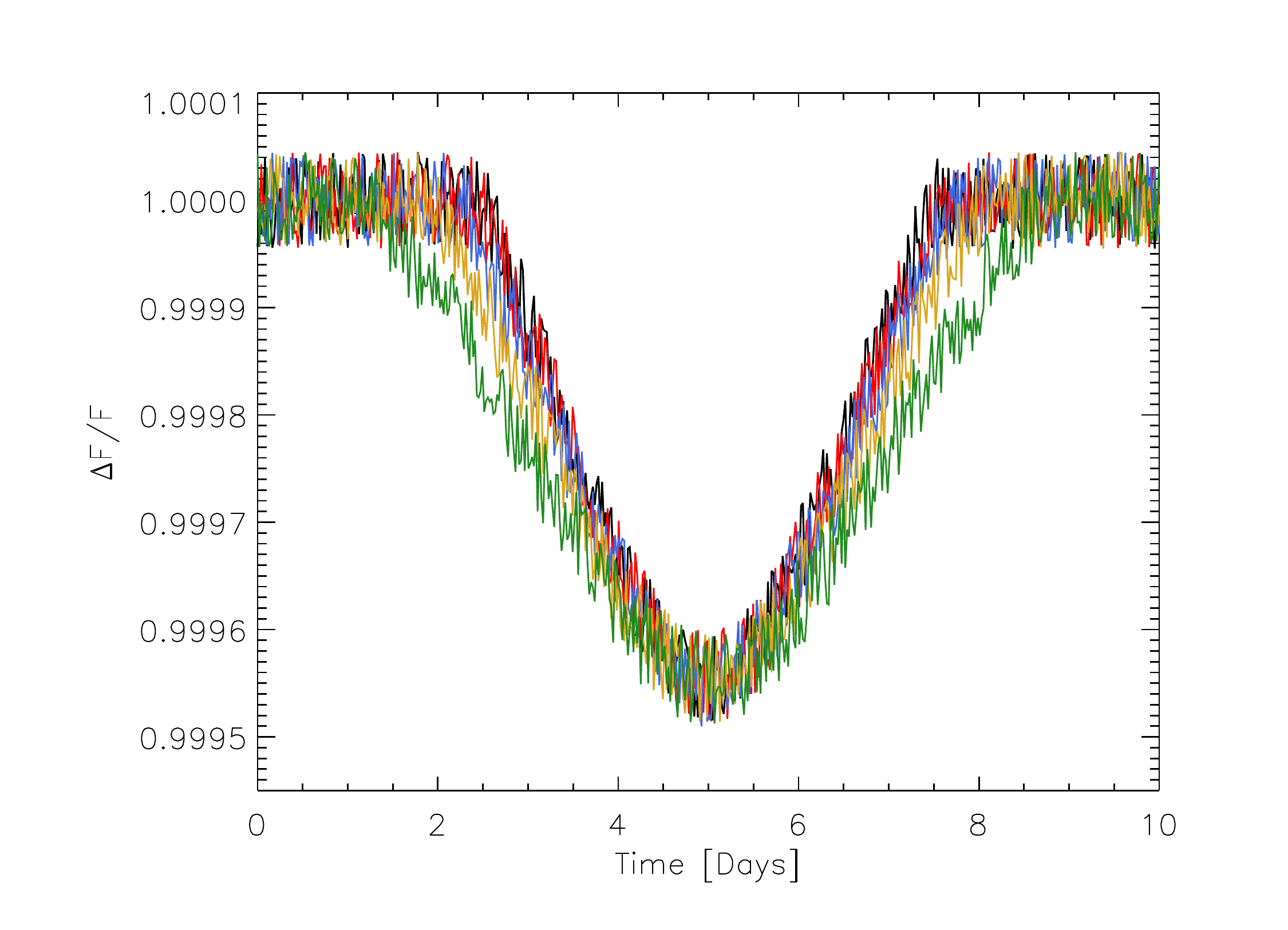}} 
\caption{Increasing photometric noise makes it more difficult to distinguish between lightcurves of different stellar inclinations. Example single spot lightcurves with varying levels of noise, for a spot located at the subobserver point on stars of different inclinations.
Black: 90$^\circ$ inclination; red: 80$^\circ$  inclination; blue: 70$^\circ$  inclination; gold: 60$^\circ$  inclination; green: 50$^\circ$  inclination. Random noise has been added to each of the lightcurves to demonstrate the blurring of the ingress as a function of noise, with \protect\subref{ingress:my-a} being perfect, noiseless data,  \protect\subref{ingress:my-b} having 25 ppm noise, \protect\subref{ingress:my-c} having 35 ppm noise, and \protect\subref{ingress:my-d} having 45 ppm noise. Although it is possible to distinguish between stellar inclinations that are very different (i.e. 90$^\circ$ and 50$^\circ$), it is more difficult to distinguish between inclinations that are within 10-20$^\circ$  of one another (e.g. 90$^\circ$ and 70$^\circ$), especially as the noise in the lightcurve increases.}
\end{figure*}

The degeneracy between stellar inclination and spot latitude plagues all spot modeling efforts \citep[e.g.][]{1996ApJ...473..388E,1997A&A...323..801K}, often making it difficult to find a unique solution that describes both inclination and spot latitude in the absence of independent constraints \citep[e.g. from a spectroscopic measurement of {\em vsini}, or by pinning the spot location(s) by planets fortuitously occulting the spot(s);][]{2010A&A...510A..25S, 2011ApJ...740L..10N}. Inclination and spot latitude affect the lightcurve by their influence on the projected size of the spot (and therefore the ultimate depth of the spot features), and also by their effect on the shape of the lightcurve during spot ingress and egress. While the depth of a spot feature is also tied to the inferred contrast against the stellar photosphere (or equivalently, the relative temperatures between the spot and the photosphere), the shape or ``shoulder'' of the spot ingress and egress in the case of ``perfect'' (noiseless) data is theoretically a fair discriminant between various inclinations. In Figure 2, we show the phase folded lightcurves of five single spot models. All five cases have spots of the same physical size (4$^\circ$ radius) that pass directly under the subobserver point (where stellar inclination plus spot
latitude equals 90 degrees, thus giving them the same projected size and thus maximum depth). In Panel \ref{ingress:my-a} we show noiseless model curves, for which the effect of stellar inclination is quite obvious-- for lower (more pole-on) inclinations, the ingress and egress of the spot is more gradual, and thus gives the feature in the lightcurve more sloping shoulders. In the subsequent panels of Figure 2, we show these same lightcurve models with 25 ppm (panel \ref{ingress:my-b}), 35 ppm (panel \ref{ingress:my-c}), and 45 ppm of noise (panel \ref{ingress:my-d}). In the presence of increasing noise, it becomes more difficult to distinguish between models of different inclination based on the shape of the lightcurve ingress and egress. However, this is mostly problematic for inclinations that are within 10-20$^\circ$ of one another, so despite this near-degeneracy, precise photometry from Kepler may be sufficient to measure inclination in the case of a single ``spot'' whose light curve displays periods of a flat continuum.

\begin{figure}
\begin{center}
\includegraphics[width=0.5\textwidth]{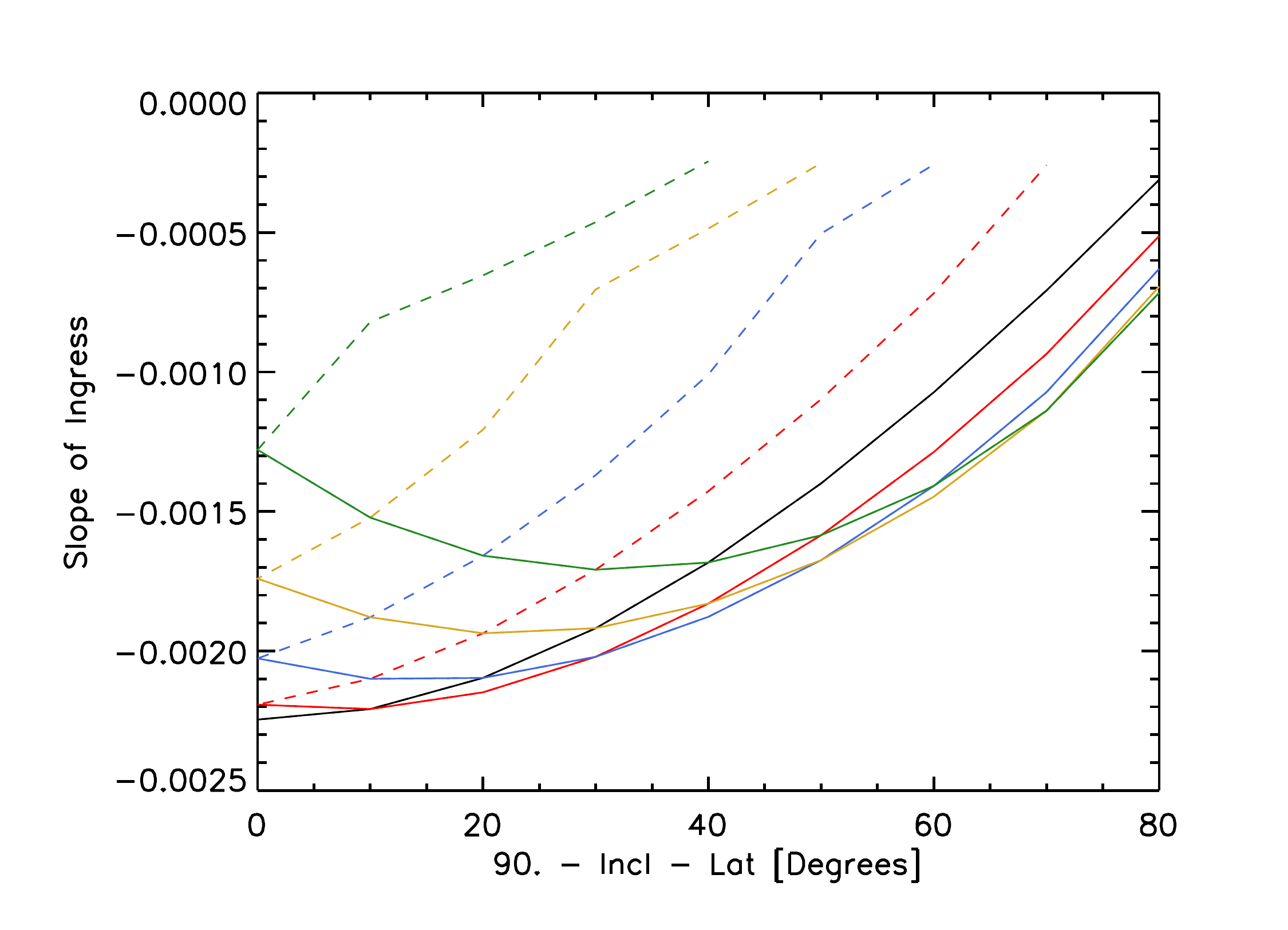}
\end{center}
\caption{The slope of the spot ingress is affected by the latitude of the spot on the star and the stellar inclination. Here we explore whether measuring the slope of the spot feature during ingress can yield information on the spot latitude and stellar inclination, by plotting the slope of a linear fit to the ingress of the spot feature versus the absolute value of its angular distance from the subobserver point (90 - Inclination - Spot latitude). The slope of ingress alone is unfortunately not sufficient to break the degeneracy between stellar inclination and spot latitude.  Colors correspond to stellar inclinations of 90$^\circ$ (black), 80$^\circ$ (red), 70$^\circ$ (blue), 60$^\circ$ (gold) and 50$^\circ$ (green). Solid lines denote spots in the northern hemisphere of the star, while dashed lines denote spots in the southern hemisphere. The y-axis is the fractional change
in flux per unit change in rotational phase (fraction of the full rotation period). The black 90$^\circ$ inclination locus describes a single line because the equator-on view makes the northern and southern hemispheres of the star symmetric, whereas for lower inclinations spots in the southern latitudes affect the lightcurve differently than their northern hemisphere counterparts. The light curves illustrated in Figure 2 correspond to points in this figure where the curves intersect the y-axis.}
\label{slpvsang}
\end{figure}

We investigated whether it would be possible to uniquely relate the shape of the ingress to some combination of latitude and longitude. We computed the slope of the ingress by fitting a straight line to the points between the beginning of the spot ingress (where the lightcurve dips below 1., or the unspotted continuum) and the half maximum depth of the feature ingress. We attempted to relate the slope of the fit to the angle between the inclination and the latitude.  In Figure \ref{slpvsang}, we show the angle between the inclination and latitude versus the slope for the subset of single spot lightcurves having a 2$^\circ$ spot. Unfortunately, a given slope intersects multiple possible combinations of latitude and inclination, thus making this plot functional only as a way of visualizing the degeneracy, rather than breaking it. 

\begin{figure}
\begin{center}
\includegraphics[width=0.5\textwidth]{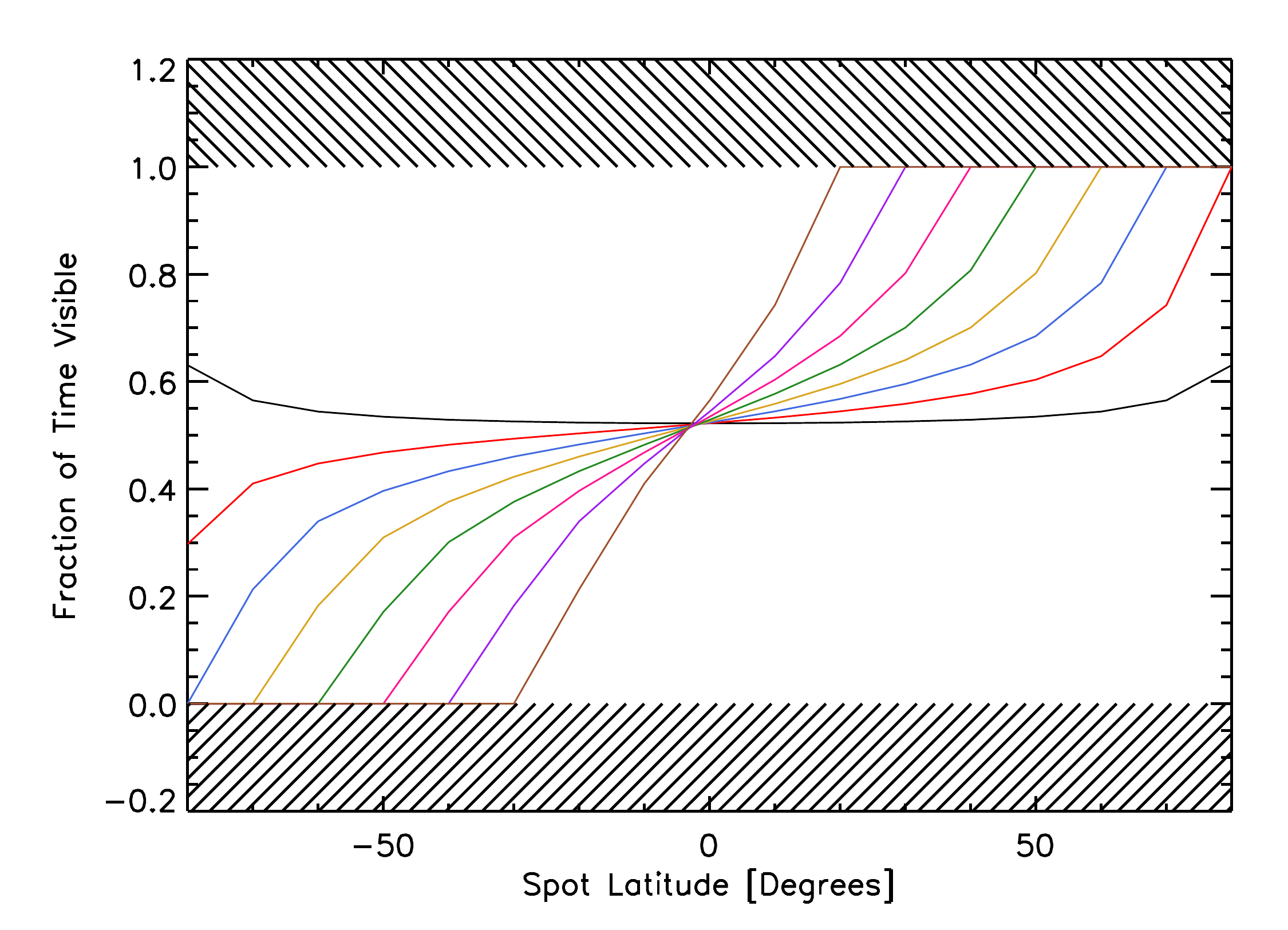}
\end{center}
\caption{The fraction of time a spot is visible is a good discriminant of spot hemisphere, but it does not
break the degeneracy between stellar inclination and spot latitude. Here we plot the fraction of time the spot is visible versus its latitude for inclinations between 20$^\circ$ and 90$^\circ$ in increments of 10$^\circ$, with colors coded as follows: black, i = 90$^\circ$; red, i =  80$^\circ$; blue, i =  70$^\circ$; gold, i =  60$^\circ$; green, i =  50$^\circ$; pink, i =  40$^\circ$; purple, i =  30$^\circ$; brown, i =  20$^\circ$. Hatching above 1 and below 0 indicates where the spot is always visible or never visible, respectively. Although measuring the fraction of time the spot is visible does not yield a unique solution for the spot latitude and stellar inclination, it does yield information on which hemisphere the spot resides in if the stellar inclination is less than 90$^\circ$-- spots in the southern hemisphere are always visible for less than half the time, while spots in the northern hemisphere are visible at least half or more of the time. }
\label{fractm}
\end{figure}

The latitude and inclination also affect how long in every rotation a given spot is visible-- for example, if one observes a star that has an inclination of 90 degrees and a single equatorial spot, that spot will be visible for exactly half of the stellar rotation, whereas a star that has a polar spot and has a lower inclination will create a lightcurve where the spot is visible most (or all) of the time. In Figure \ref{fractm}, we show the fraction of time a given feature is visible versus the spot latitude for the subset of lightcurves having 4$^\circ$ spots and  inclinations between 20$^\circ$ and 90$^\circ$ in increments of 10$^\circ$. Hatched regions of the plot (above 1 and below 0) indicate where the spot is always visible, or never visible, respectively). Again, this plot elucidates that there are many combinations of latitude and inclination that result in a given fraction of time visible. However, it also points out that with the exception of a star that is exactly equator-on, the fraction of time visible is helpful for discriminating between which hemisphere a given spot is location in; spots that are visible less than half of the time are located in the southern hemisphere (at negative latitudes), while spots that are visible more than half the time must be located in the northern hemisphere (at positive latitudes). Knowing the fraction of time the spot was visible would allow one to decide whether the solid or dashed set of branches in Figure 3 were the appropriate set, but it would still not break the degeneracy between them. 

\begin{figure}
\begin{center}
\includegraphics[width=0.5\textwidth]{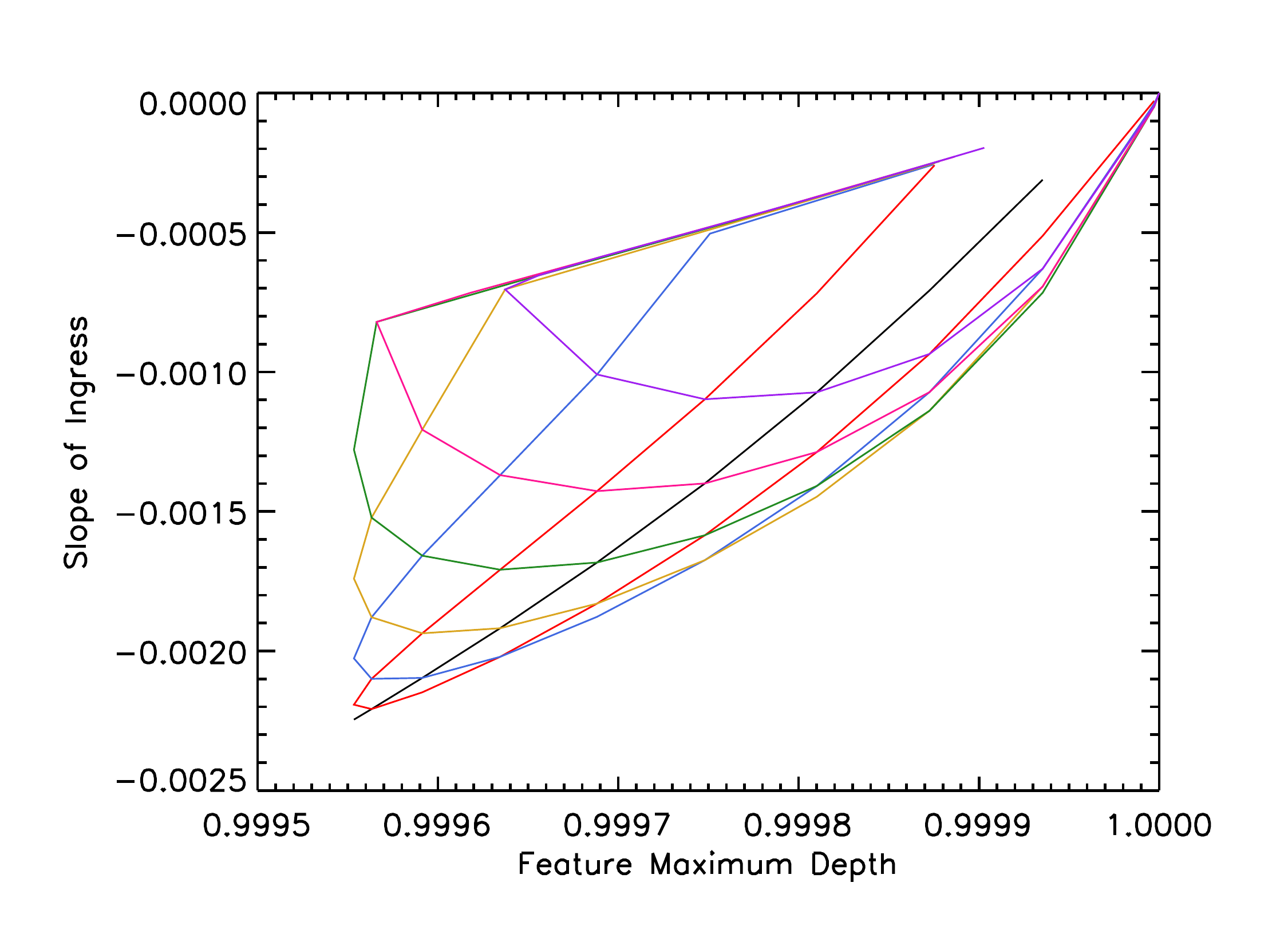}
\end{center}
\caption{Lightcurve feature depth
and ingress slope do not break the degeneracy between stellar inclination
and spot latitude.  The slope of the spot ingress and the maximum depth of the spot feature are both observables, related to the stellar inclination, spot latitude and spot size. Here we show these quantities for models with 2$^\circ$ spots and inclinations between 20$^\circ$ and 90$^\circ$ in increments of 10$^\circ$, with colors coded as follows: black, i = 90$^\circ$; red, i =  80$^\circ$; blue, i =  70$^\circ$; gold, i =  60$^\circ$; green, i =  50$^\circ$; pink, i =  40$^\circ$; purple, i =  30$^\circ$. The y-axis is the fractional change
in flux per unit change in rotational phase (fraction of the full rotation
period). The measurement of these quantities does not yield a unique combination of spot latitude and stellar inclination, but for some regions of parameter space one can narrow the latitude and inclination down to a family of possible solutions.}
\label{depthvsslp}
\end{figure}

The depth of the feature and slope of the ingress and egress are the most easily observable features of a lightcurve, and they are both influenced by the spot latitude and stellar inclination.  In Figure \ref{depthvsslp} we show the relationship between the depth of the spot feature and the slope of ingress (as described earlier) for the subset of models having spots of 2$^\circ$ for models with inclinations between 90$^\circ$ and 30$^\circ$. Clearly the stellar inclination creates unworkable degeneracy between these two quantities. Conversely, this means that an independent determination of the inclination is extremely helpful.

The systematic characterization of these various disheartening degeneracies clearly demonstrates that in these very simple, single-spot cases, there are many sets of model parameters that produce extremely similar lightcurves. In the following section, we explore whether the presence of additional spots helps to break or worsen this degeneracy.

\subsection{Period Recovery}
\label{sec:perrecover}

As mentioned in Section \ref{sec:intro}, ``the'' stellar rotation period as traced by starspots may be something of a misnomer-- in the case of a differentially rotating star, the distribution of rotation periods derived from periodogram analysis of a given lightcurve represents not a single period, but the periods of all spots at their particular latitudes. Indeed, one of the primary motivations for spot modeling is to deduce what the distribution of spot latitudes and their corresponding periods, such that the sense and magnitude of the differential rotation may be derived. To complicate matters, the peaks in the computed periodogram do not necessarily correspond to true periods in the data: true periods are present amidst aliases, harmonics and blends, so to simply take the periodogram peak with the most power as the true period may be dangerous. Understanding the effect of the spot distribution on the periods derived is important, as the astrophysical conclusions drawn may rely on it (e.g., two similarly-sized spots located on opposite longitudinal hemispheres will modulate the lightcurve at half the true period of the star, which may then result in an incorrectly inferred rotation period and thus incorrect stellar age). 

In this section, we use our set of two-spot synthetic lightcurves to explore the influence on the spot parameters on period recovery using periodogram analysis via the prescription of \citet{1986ApJ...302..757H}, with a focus on how spot parameters affect the derivation of the true period (or periods) present in the data \citep[for an exploration of the effects of time sampling, see][]{1993A&A...269..351L}. For example, one key question is how often one recovers one or more periods actually present in the model, versus how often one recovers a harmonic of the actual period or a mixture of multiple periods present in the lightcurve (in the case of models with differential rotation). Which physical situations lead to the measurement of aliases of the rotation period, rather than the period itself? In the case of differential rotation, what parameters influence the dominant period found?

\begin{figure*}
\begin{center}
\subfloat[][]{\label{londiffvper:p-a}\includegraphics[width=0.5\textwidth]{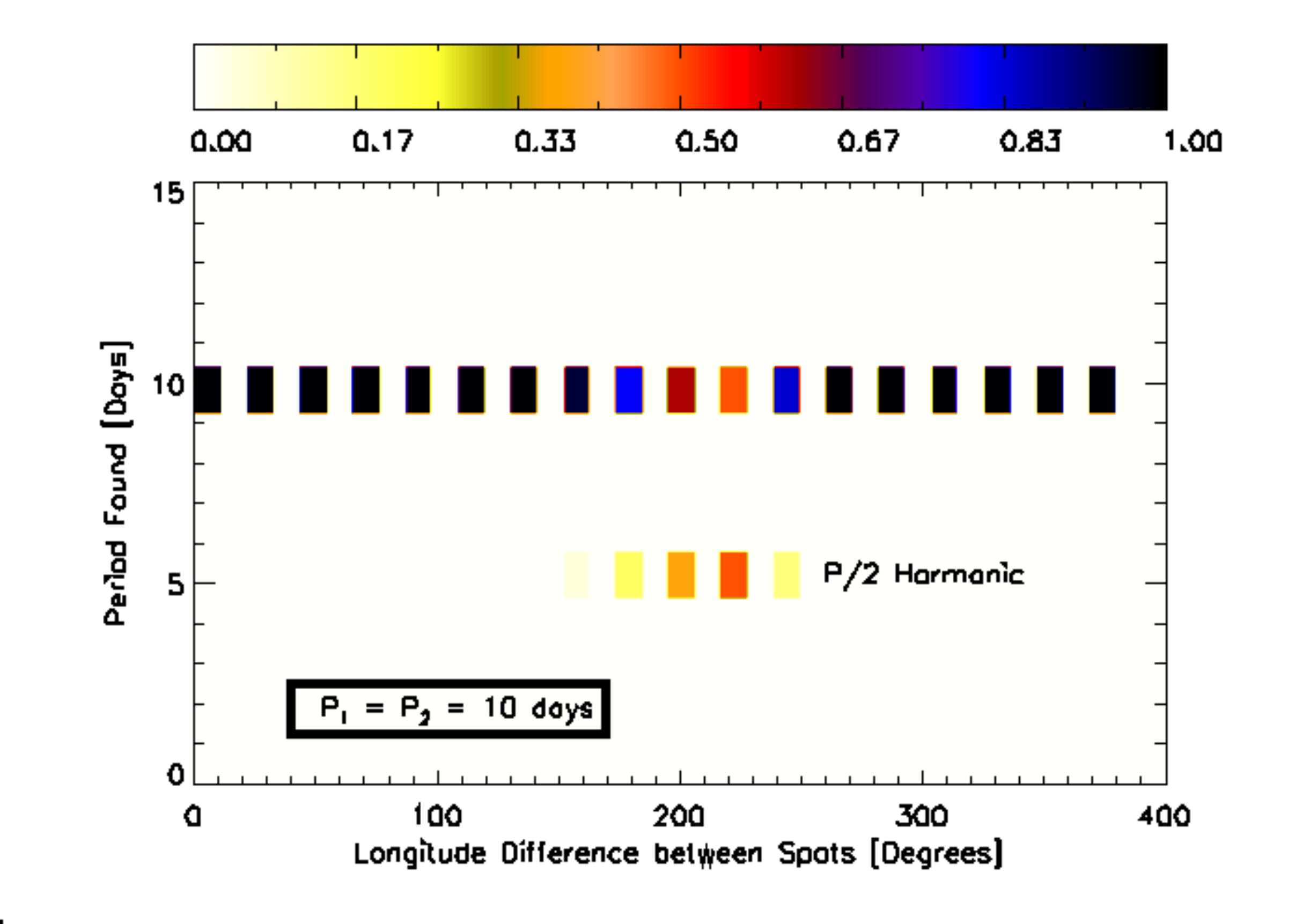}} 
\subfloat[][]{\label{londiffvper:p-b}\includegraphics[width=0.5\textwidth]{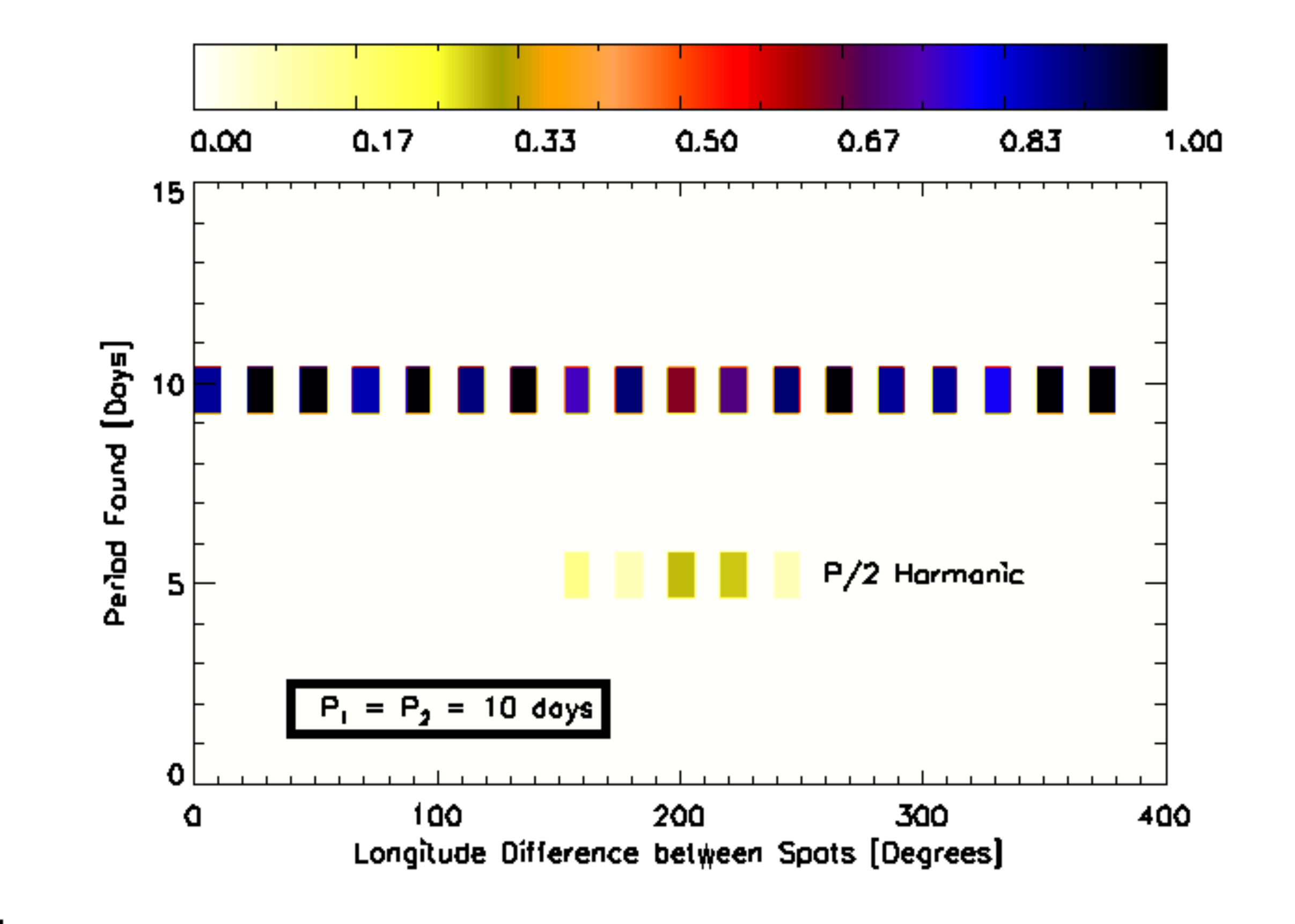}} \\
\subfloat[][]{\label{londiffvper:p-c}\includegraphics[width=0.5\textwidth]{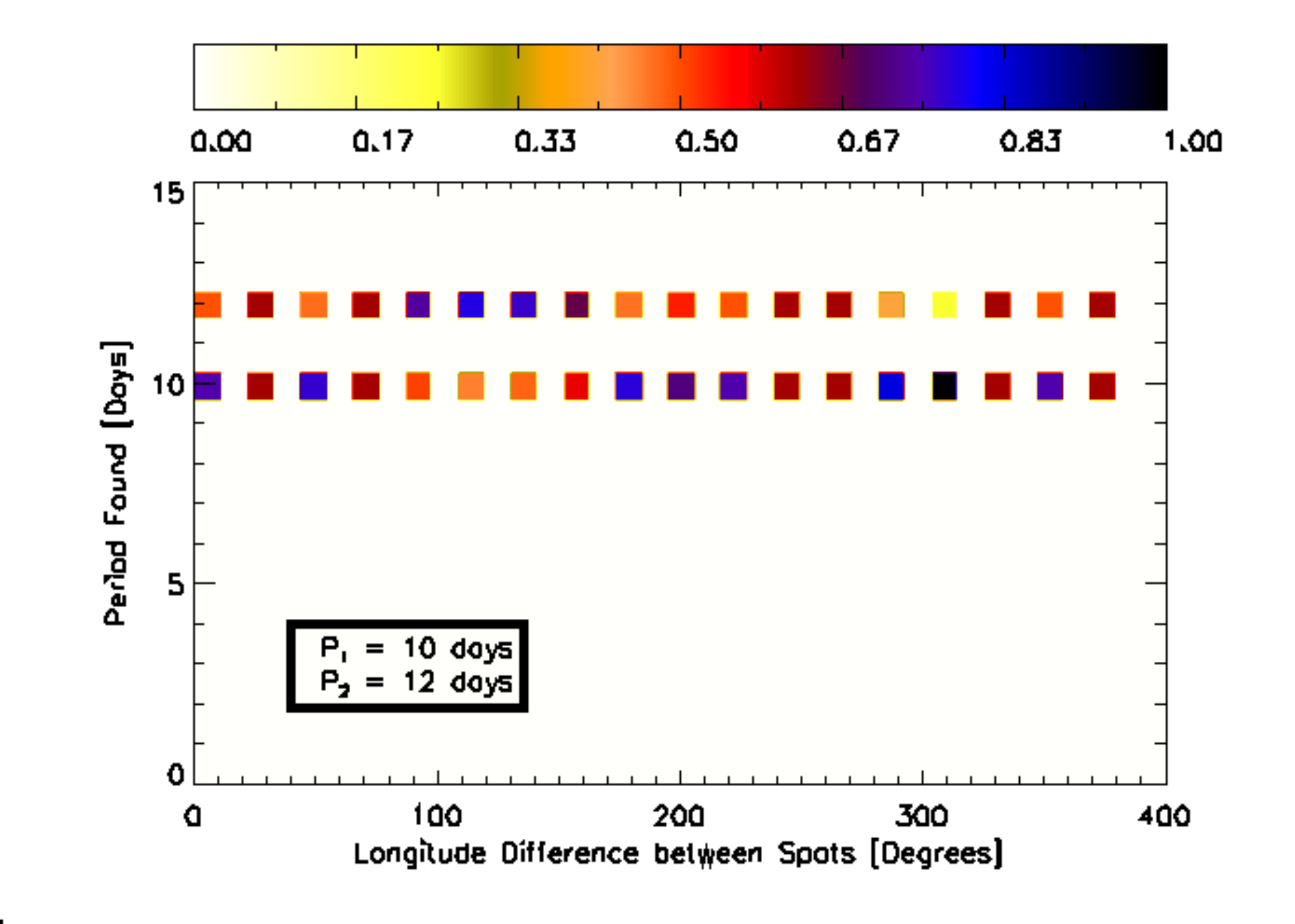}} 
\subfloat[][]{\label{londiffvper:p-d}\includegraphics[width=0.5\textwidth]{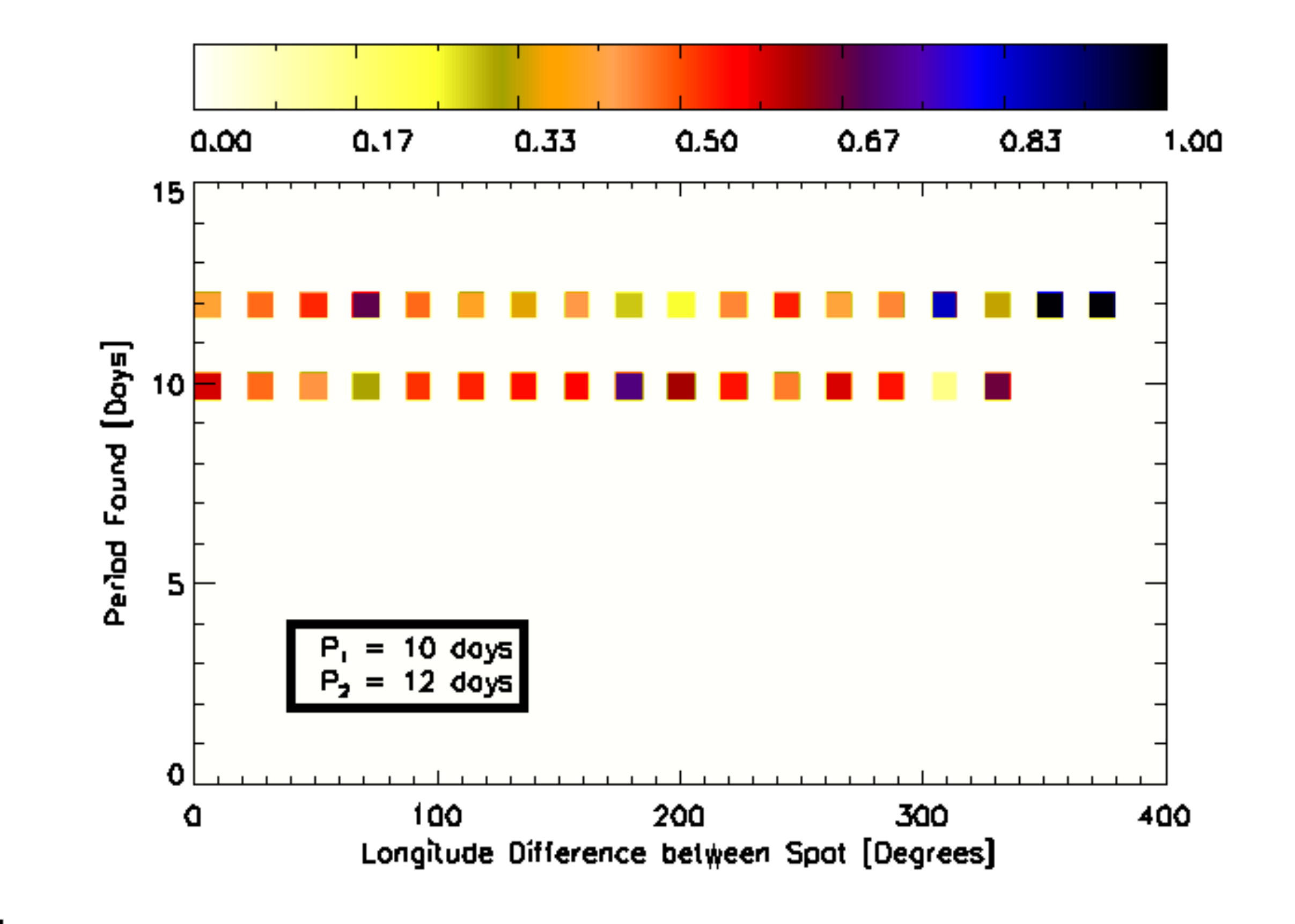}} 
\end{center}
\caption{Stars with spots at roughly opposite longitudes may appear to have a period of half their true period, due to the symmetry of their spot distribution-- however, the presence of differential rotation helps mitigate the chances of misinterpreting the period of the lightcurve due to symmetry in the spot distribution. Here we show period found as a function of longitude separation of the two spots for the four sets of lightcurves: (a) 90 degree inclination, no differential rotation; (b) 45 degree inclination, no differential rotation; (c) 90 degree inclination with differential rotation; and (d) 45 degree inclination with differential rotation. The color bar shows the fraction of cases that appear in each bin-- i.e. a value of 1 and complete black indicates that 100$\%$ of cases landed in that particular bin. In panels A and B, the true period of the lightcurves is 10 days, so if the period was always found correctly, filled boxes would only appear at 10 days. In cases where the spots are separated by between 140 and 220 degrees and are of equal sizes, however, a harmonic at half of the true period is sometimes found. Panels C and D show the periods found for lightcurves with differential rotation (true spot periods are 10 days and 12 days). Here we can see that there are no cases for which P/2 harmonics of the true period are found (i.e. there are no 5 or 6 day periods found), and no spurious periods.}
\label{londiff}
\end{figure*}

We begin with the simplest set of lightcurves: two spots (of a range in size, described in Section 3.1) on a 90$^\circ$ inclination star with no differential rotation. Here the northern and southern hemispheres are symmetric from the point of view of the observer and only one possible period exists to be recovered (chosen to be 10 days).  Figure \ref{londiffvper:p-a} shows the period recovery as a function of the longitude separation between the two spots in the model for this solid-body rotation, 90$^\circ$ inclination set.  In almost all cases, the actual period was recovered, but there was a small subset of lightcurves which were prone to finding the P/2 alias of the true period. These cases were all where the two spots were of equal size and were between 140$^\circ$ and 220$^\circ$ apart in longitude, so that two identical spots on opposite sides of the
star look similar to one spot on a star rotating twice as fast. For all longitude separations greater than 220$^\circ$ or less than 140$^\circ$, 100$\%$ of the lightcurve periods were found correctly, but in the cases where the two spots where separated by between 140 and 220 degrees, the strongest peak in the periodogram was found at a period of half its true value. Stars with such a spot distribution do exist in nature-- \citet{2009MNRAS.400..451C} noted that in observations of the Coma Berenices open cluster taken during two separate observing periods, the lightcurves of several members yielded periods at half of their expected value from the cluster rotation-color relationship. The true periods of these stars likely lie along the same rotation-color relation as the rest of the cluster, but symmetric spot groups on opposing hemispheres make the period seem to be half its true value. For such spot configurations, even the most precise data may not be able to definitively state the true rotation/age of the star, unless spot evolution breaks this symmetry during the course of monitoring.

In the case of our set of model lightcurves that had two spots, no differential rotation, but a 45 degree inclination, we find a very similar result (shown in Figure \ref{londiffvper:p-b}). Again, the correct period is recovered for almost all lightcurves, with the exception of those with spots that were the same size and spaced between 140$^\circ$ and 220$^\circ$ apart. For spots outside this longitude separation range, boxes that have a value less than 1 (i.e. are not filled completely black) are due to some lightcurves having spots that are either hidden or always in view, thus resulting in no period being found. 

Figures \ref{londiffvper:p-c} and \ref{londiffvper:p-d} are analogous to the results described above, but for the set of lightcurve models with 90 and 45 degrees inclination (respectively) which included differential rotation, such that both a 10 day and 12 day period were extant in the model. Interestingly, the dominant period found in these cases is {\em never} a harmonic of one of the two true periods present in the data, so differential rotation mitigates the issue of harmonics posing as the true rotation period in the data. Differential rotation forces the longitude difference between the two spots to change with time, eventually leaving the 140-220 degrees range, at which point the light curve cannot be reproduced by a single spot on a star with half the period.

\begin{figure} 
\begin{center}
\subfloat[][]{\label{subfig:my-b}\includegraphics[width=0.5\textwidth]{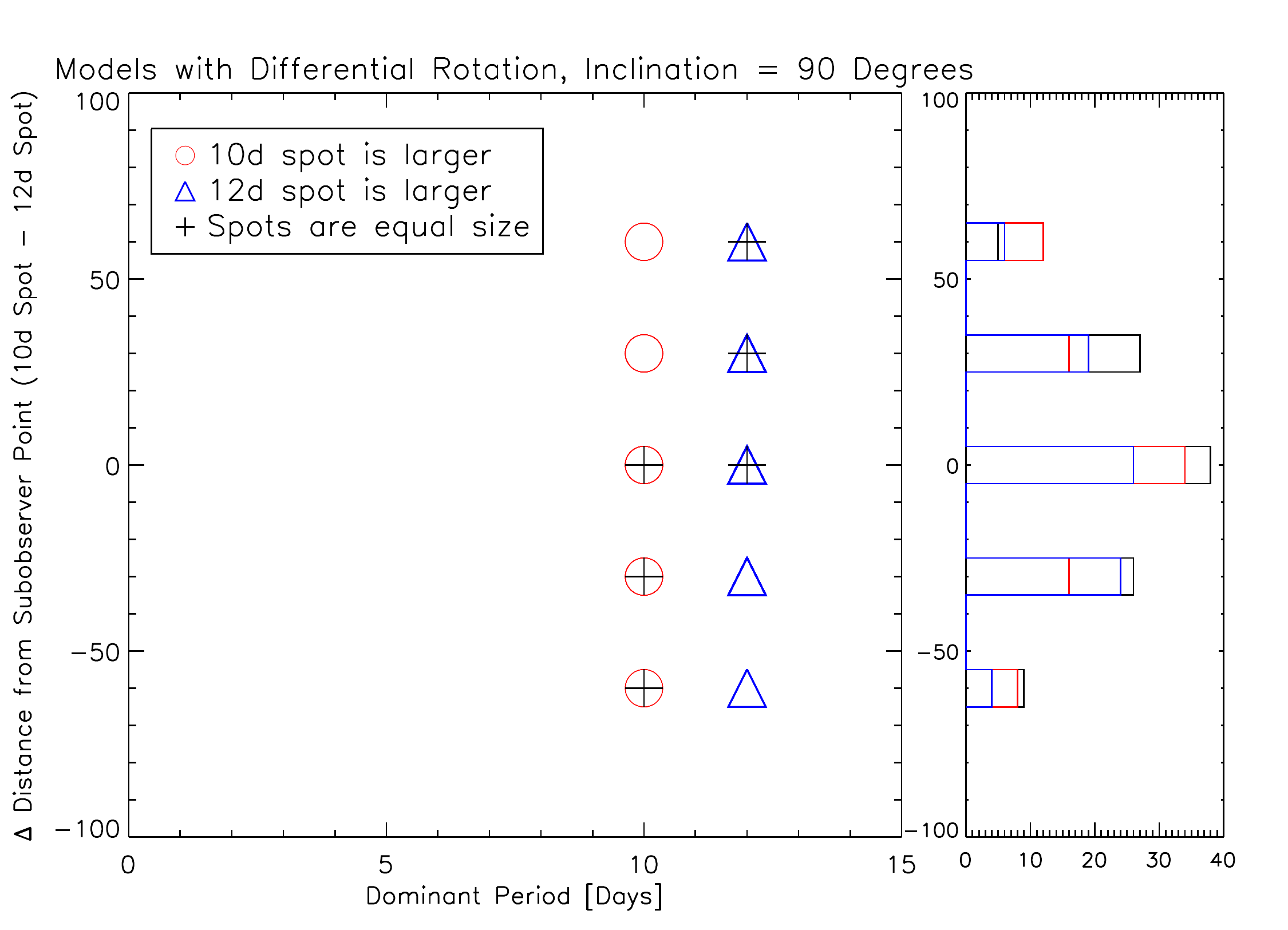}} \\
\subfloat[][]{\label{subfig:my-a}\includegraphics[width=0.5\textwidth]{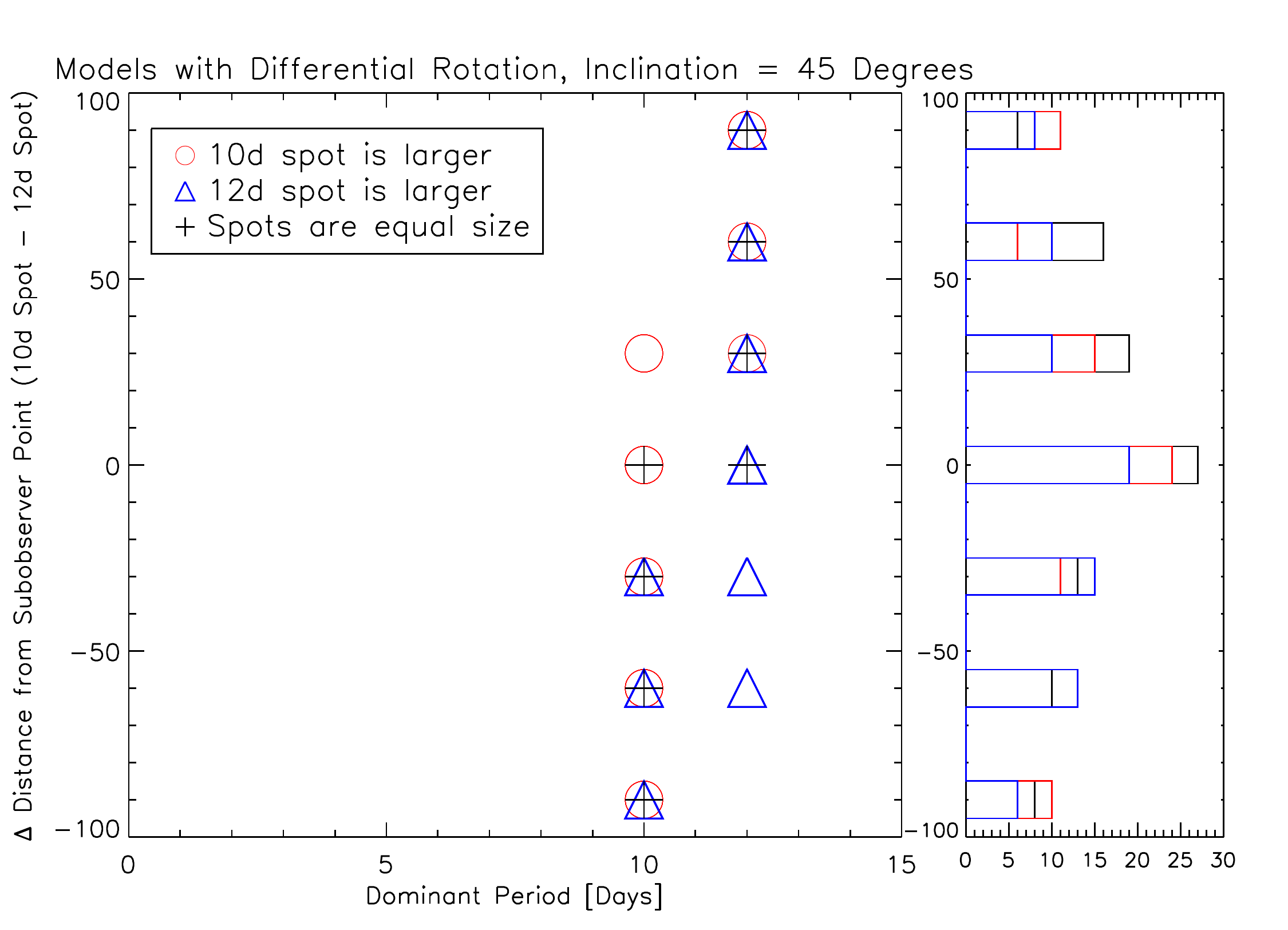}} 
\end{center}
\caption{The dominant period found tends to be that of the spot with the largest projected size. In models with two spots and differential rotation (such that one spot has a 10 day period and the other spot has a 12 day period), the period found is a function of the projected size of the spots (their physical size combined with their location on then star). Here we show the difference in distance from the subobserver point for the two spots versus the dominant period in the periodogram for 90 degree inclination at top and 45 degree inclination at bottom. The histograms at right show the number of lightcurves that fell into each bin. In the top panel, where the stellar inclination is 90 degrees, the period found is always that of the larger spot. In the cases where the spots are equal in size, the distance to the subobserver point (and thus the projected size of the spot) becomes important, and the period found is that of the spot with the larger projected size. } 
\label{distdiff_v_period_dr90_simple_hist} 
\end{figure} 

For the set of lightcurves with 90 degree inclination and differential rotation, the period found is always that of the larger of the two spots. This effect can be seen in Figure \ref{distdiff_v_period_dr90_simple_hist}a, where we show the period found versus the difference in distance from the subobserver point for the 10 day period spot compared to the 12 day period spot. When the 10 day period spot is larger than the 12 day spot, the dominant period is 10 days (red circles), and similarly when the reverse is true, the period found is 12 days (blue triangle). The distance to the subobserver point is only of consequence when the spots are equal in size (black crosses), in which case the period found is that of the spot which is closest to the subobserver point (and therefore has the largest projected area). 

The results are somewhat more complex for the set of lightcurves with inclination of 45 degrees that include differential rotation, shown in Figure \ref{distdiff_v_period_dr90_simple_hist}b. Because the viewing angle of the spot is no longer symmetric with respect to the observer, there is now a larger variety of possible subobserver distances. For the simplest case the result is still clear cut: When the spots have equal size (black crosses), the dominant period found is that of whichever spot is closer to the subobserver point (and thus has the largest projected area). However, the larger range of possible subobserver distance now competes more with the physical spot size for dominance of the lightcurve-- in the two cases where the larger spot is further away from the subobserver point, one may find the period of either spot. In fact, the period found is that of the spot with the largest projected area. 

\begin{figure}
\begin{center}
\subfloat[][]{\label{perdiff-a}\includegraphics[width=0.5\textwidth]{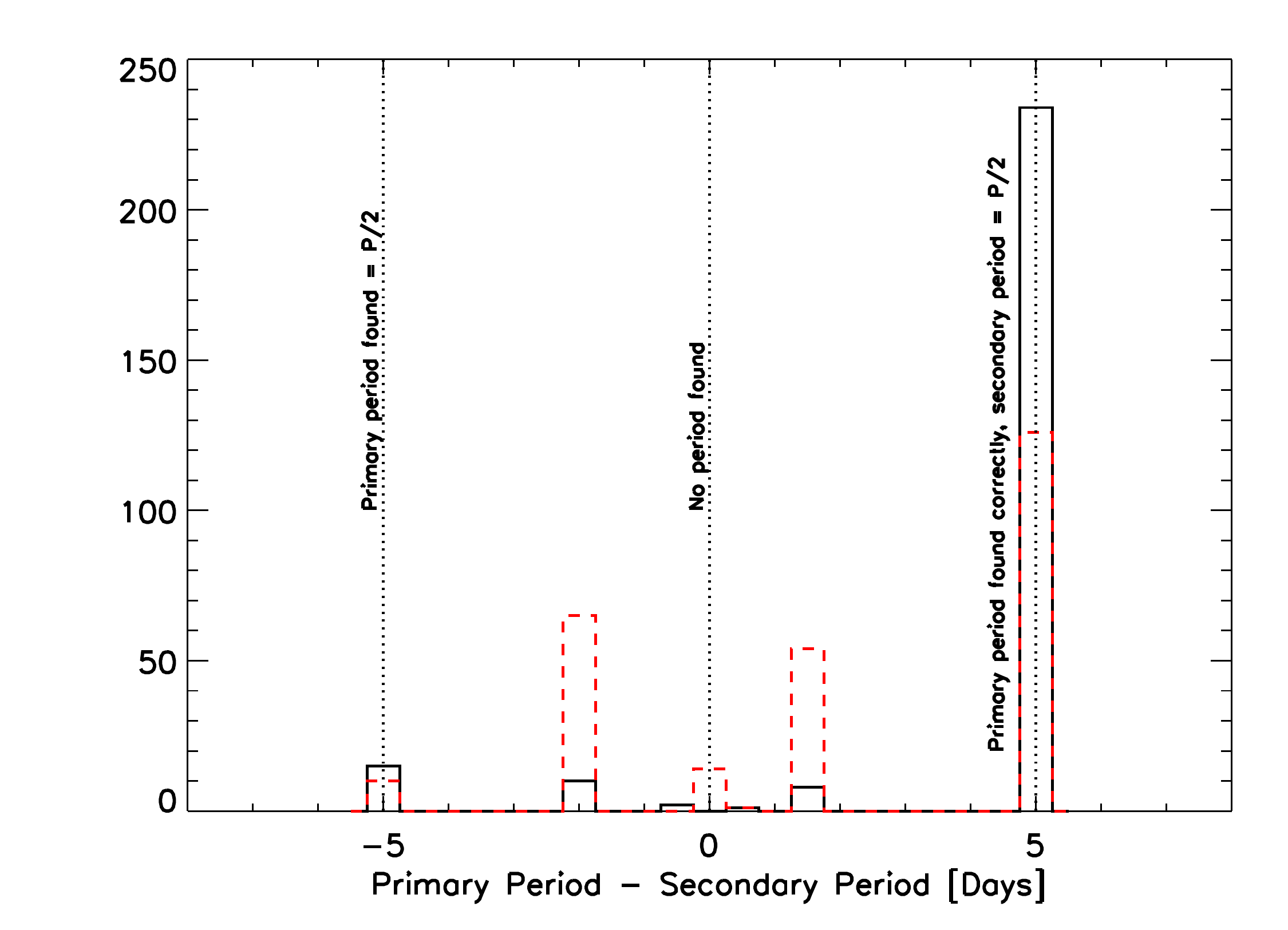}} \\
\subfloat[][]{\label{perdiff-b}\includegraphics[width=0.5\textwidth]{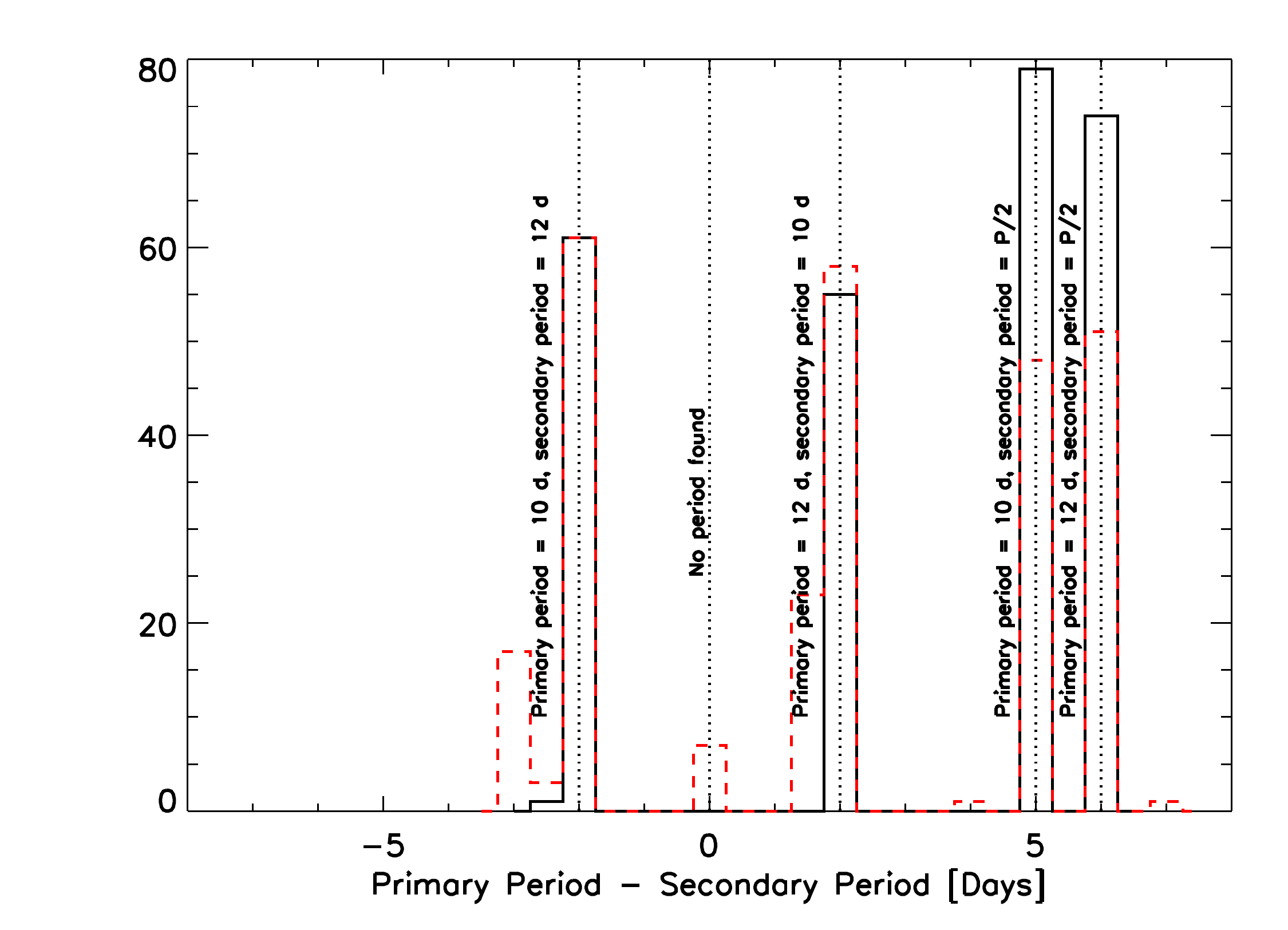}} 
\end{center}
\caption{The strongest peaks in a given periodogram may not characterize all periods present. Here we show histograms of the dominant period minus the second most significant period found, for cases without differential rotation (top) and with differential rotation (bottom). Black solid lines indicate models with a stellar inclination of 90 degrees, dashed red lines indicate models with a stellar inclination of 45 degrees. In the top panel, the correct 10 day period is found as the primary period for most models, though in some cases the P/2 harmonic may be found instead. When the P/2 harmonic (5 days) is found as the dominant period, the true period often appears as the second strongest peak in the periodogram. In the bottom panel, we see that in many cases, the two most prominent peaks in the periodogram are representative of the true periods in the model: a spot at 10 days and a spot at 12 days. However, it is actually more common that the second strongest peak is the P/2 harmonic of the primary period found, rather than being representative of the other true period in the data.  }
\label{perdiff}
\end{figure}

We also investigated whether the second most significant peak in each periodogram was found at the period of the second spot, or whether a harmonic of the dominant period was more likely to be found. Figure \ref{perdiff} shows the difference in period  between the dominant period and the second most significant peak in the periodogram for two-spot models with no differential rotation (panel \ref{perdiff-a}) and with differential rotation (panel \ref{perdiff-b}). In Figure \ref{perdiff-a}, most of the lightcurves lie at P$_{dom}$ $-$ P$_{2}$ = 5, indicating that the primary period found was correct (10 days), and that the second highest peak was found at P/2 (5 days). For a small number of cases we see that P$_{dom}$ $-$ P$_{2}$ = $-$5, or that the harmonic P/2 was the dominant period, and that the true period of 10 days had the second highest significance. Figure \ref{perdiff-b} shows the difference in primary and secondary periods for lightcurves with differential rotation. Here there are substantial number of cases both at $-$2 and 2, the former indicating that the primary period found was 10 days and the second most significant peak was found at 12 days (and vice versa in the latter case of $+$2). In these cases, the two strongest peaks found were representative of the true periods of the two spots in the model. There are also a substantial number of cases which lie at P$_{dom}$ $-$ P$_{2}$ = 5 or 6; in these cases, the primary period found was 10 or 12 days, respectively, and the secondary period found was at P/2. In both the cases with differential rotation or without, for the most part only the true periods or harmonics are populated in the histograms (particularly for the 90$^\circ$ inclination models, black solid line). However, there are a few other bins found$-$ these are cases where the secondary peak in the periodogram represented neither a true spot period nor a harmonic, but rather power elsewhere in the periodogram. These tend to be cases where the next strongest power lay close to the true period, but centered a day or so off of the true peak.  

\begin{figure}
\begin{center}
\includegraphics[width=0.5\textwidth]{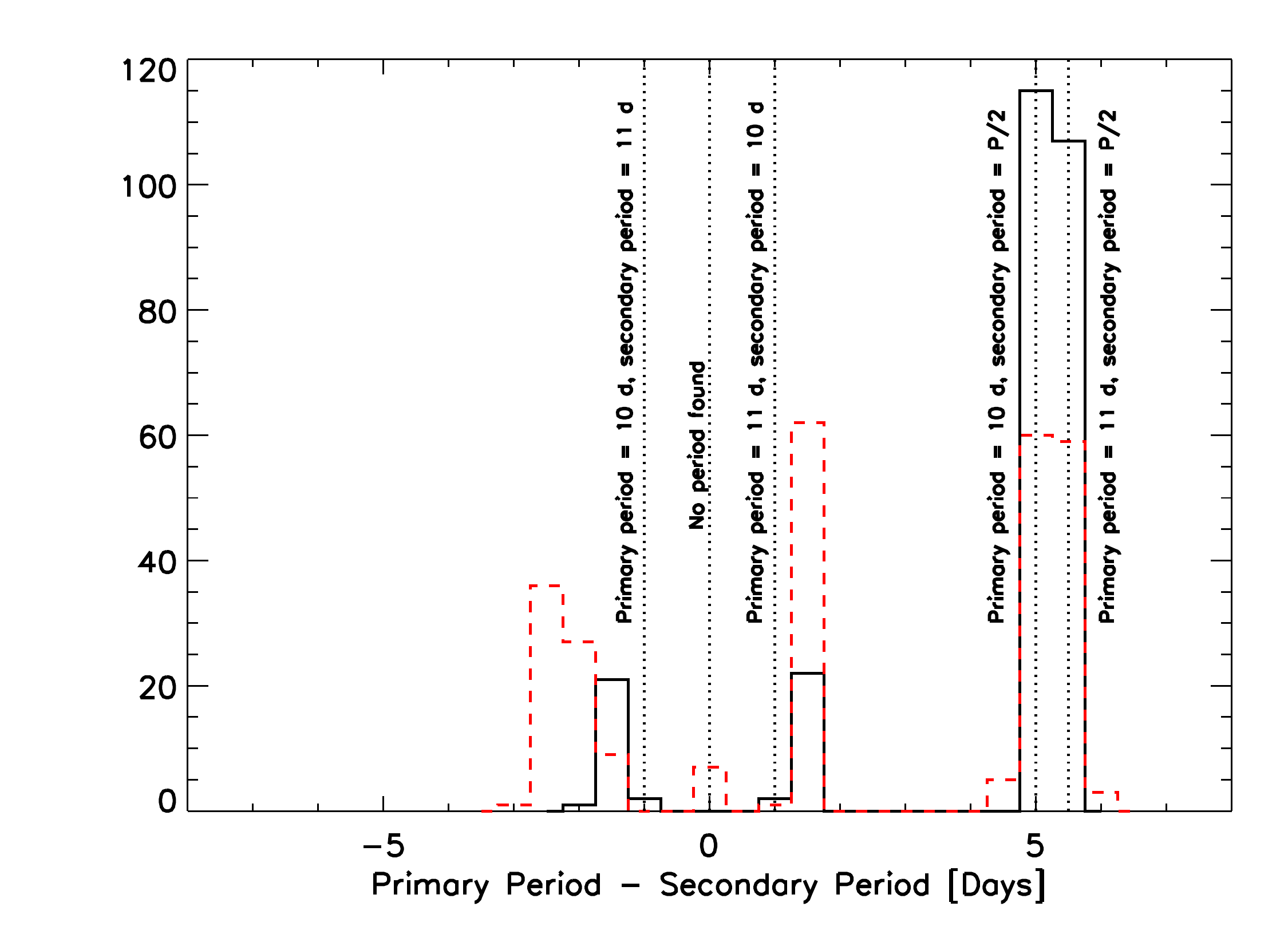}
\end{center}
\caption{Recovering the true value of differential rotation from the periodograms alone is subject to a great deal of uncertainty, especially when the amount of differential rotation is relatively small. Here we show histograms of the dominant period minus the second most significant period found for lightcurves with a lesser amount of differential rotation (spots at 10 and 11 days, as opposed to 10 and 12 days in all cases previously discussed). Here we see that again, the two strongest peaks in the periodogram are more often one of the periods in the model and its P/2 harmonic, and are therefore not necessarily representative of all the periods present. In addition, for these models with less differential rotation, the fact that the periods present in the model are closer together (10 and 11 days) leads to some confusion-- it is really very uncommon that the two strongest peaks found represent the two periods present. More often, one gets a second period that is a bit off from the true period (e.g. if the 10 day period is the strongest peak in the periodogram, the second peak is sometimes 12 days, not 11 days).}
\label{perdiffless}
\end{figure}

Finally, we take a brief look at the effect of the {\em amount} of differential rotation on the period recovery. The particular value we have chosen for the differential rotation in the models discussed thus far (20$\%$) is at the high end of the range of expected differential rotation for typical solar-type stars. We therefore computed an additional set of lightcurves with only 10$\%$ difference in periods (i.e. two-spot models with spot at a 10 day period and a spot with an 11 day period). In Figure \ref{perdiffless} we show the primary and secondary periods found for these lightcurves. Interestingly, in these cases it was far more common to recover the P/2 harmonic as the second most significant peak than to find a peak at the true period of the second spot. The reason one rarely finds the second peak at the other truly representative period is that for our chosen time series length of 90 days (approximately the length of a single quarter of Kepler data) the frequency resolution (1/T, or 0.011 d$^{-1}$) is similar to the frequency separation between the two rotation periods (0.0091 d$^{-1}$). For these closely spaced periods over the duration of observing, the differential rotation appears as asymmetric lobes on the peak of the dominant period, rather than independent, well-separated peaks in their own right. These lobes may not be found in an automated search for periodogram maxima of the sort we have considered here, but one could potentially consider fitting multiple components to the periodogram distribution, or iteratively removing the larger periodic signals in the data to recover their locations (this latter approach is essentially what is done to place spots at the appropriate periods when implementing spot modeling with Cheetah). Indeed, one of the advantages to applying a spot model to the data is that these models are capable of recovering quite small amounts of differential rotation, which wouldn't be readily detectable by periodogram analysis \citep[see for example][]{2006ApJ...648..607C}.

\section{Discussion and Future Work}

\begin{quotation}
{\em ``All models are wrong. Some models are useful.''}$-$ \citet{box87}
\end{quotation}

The modeling of stellar spots is plagued by inherent degeneracies, and so often an observed lightcurve may be closely reproduced by multiple models with different spot parameters. It is therefore of utmost importance that those who undertake the task of turning an integrated lightcurve into a spot distribution parametrize their ignorance and understand the uncertainties of a given model. While a single model of a given star may not represent ground truth, differences between many lightcurves treated in a uniform way can reveal trends and relationships that may indeed be physically motivated. In the case of the Kepler data, which provides not only precision but a very large set of stars to work with, one may hope that complementary spot modeling methods will give detailed results for individual stars as well as reveal statistically interesting relationships that take advantage of the sample size. 

In this paper we have used synthetic lightcurves generated by a simple spot model, Cheetah, to investigate whether morphological metrics of lightcurves can constrain spot parameters even before a physical model is fit to the data. We find that in general most parameters of simple spot models are indeed highly degenerate, but that in some cases the precision of Kepler does allow stellar inclination to be determined. We additionally find that the fraction of time a spot feature is visible can be used to constrain which hemisphere it appears on, and that the presence of differential rotation mitigates the chances of mistaking a P/2 harmonic for the true rotation period of the star. We have considered only what may be derived from the photometry itself; in practice it may be that additional data is available to further constrain the model parameters. Such data tends to be preferentially available for planetary system host stars, as efforts to measure the planets' masses by radial velocity variations yield spectra and therefore a measure of {\em vsini}. In addition several cases of transiting exoplanets have been observed to occult spots, or parts of spots \citep[e.g.][]{2010A&A...510A..25S, 2011ApJ...740L..10N,2011ApJ...743...61S}, which when combined with the planetary orbital geometry can constrain not only the stellar inclination but also the location of spots along active latitudes or longitudes. 

Besides precision and large sample size, the Kepler data offer exceptionally {\em long}, largely uninterrupted monitoring. The long time baseline of these data provide far more stringent constraints on the possible stellar inclination due to the emergence and dissipation of spots at varying latitudes than short lightcurves with static spots, as we have dealt with here. While beyond the scope of this particular paper, we hope that additional data will remove some of the uncertainty in the stellar inclination.  

An unsolved issue in the prior analysis is that the unspotted light level of a given star is not known. In practice, we set the brightest point in any given lightcurve as equal to ``1'', or the unspotted continuum, but in reality polar spots or spotted belts around the star may exist but have no time varying effect on the stellar intensity. In addition, it may be impossible to know the spot distribution on certain parts of a star due to geometry-- as the inclination decreases, the circumpolar region of the opposite pole moves out of the observer's sight where they cannot influence the lightcurve, so spot information in that region is lost completely. We therefore remind the reader that results from this and similar models indicate minimum relative spot coverage only. 

In the current iteration of the model, spot evolution is not included. We do plan to include spot evolution in a future version of Cheetah, as it would be potentially interesting to derive spot evolution timescales for our targets, but for now we cautiously treat the spots as fixed in time.  The distribution of differential rotation parameters is far better constrained by observation than the spot evolution timescale, which seems to vary from spots that barely survive a rotation period (as is the case on our own Sun) to very long-lived spots \citep[as seems to be the case for, e.g., Proxima Cen;][]{1998AJ....116..429B}.

We anticipate that the rate of stellar astrophysics investigations with Kepler will increase greatly in the extended mission, as the reduction pipeline (PDC-MAP) continues to be improved and the data become immediate available to the community. It is our intent to apply Cheetah to produce simple models of the spot distribution for as many stars as possible, and to eventually make the code public after validating it on real data. From the confluence of the technique presented and complementary in-depth treatment of individual stars through other means, we expect that the next several years will bring us closer to ``seeing'' stellar surfaces and the place of our own Sun among them. 

\acknowledgements
LMW and GB thank Ansgar Reiners for valuable discussions on the development of the program.

\clearpage

\end{document}